\documentclass[journal,compsoc]{IEEEtran}

\usepackage[nocompress]{cite}
\usepackage{balance}
\usepackage{amsmath}
\usepackage{centernot}

\usepackage[pdftex]{graphicx}
\graphicspath{{../pdf/}{../jpeg/}}
\DeclareGraphicsExtensions{.pdf,.jpeg,.png}
\usepackage{multirow}
\usepackage{pgfplots}
\usepackage{tikz}
\usetikzlibrary{positioning}
\usetikzlibrary{shapes.geometric}

\tikzset{
    astnode/.style={circle, draw=black, align=center, thick}
}

\usepackage{mdframed}
\mdfdefinestyle{mpdframe}{
	frametitlebackgroundcolor   =black!15,
	frametitlerule              =true,
	roundcorner                 =5pt,
	middlelinewidth             =0.8pt,
	innermargin                 =0.2cm,
	outermargin                 =0.2cm,
	innerleftmargin             =0.2cm,
	innerrightmargin            =0.2cm,
	innertopmargin              =0.2cm,
	innerbottommargin           =0.2cm
}

\usepackage[ruled,vlined,linesnumbered]{algorithm2e}
\makeatletter
\newcommand{\removelatexerror}{\let\@latex@error\@gobble}
\makeatother

\usepackage[]{xcolor}
\newcommand{\TODO}[1]{\textcolor{red}{#1}\GenericWarning{}{LaTeX Warning: TODO: #1}}\newcommand\todo\TODO

\usepackage{subcaption}

\usepackage[para,online,flushleft]{threeparttable}
\usepackage{longtable}
\usepackage{booktabs}
\usepackage{url}
\usepackage{comment}
\usepackage{hyperref}
\usepackage{xspace}
\usepackage{cleveref}
\renewcommand{\sectionautorefname}{Section}
\let\subsectionautorefname\sectionautorefname
\let\subsubsectionautorefname\sectionautorefname 
\renewcommand{\chapterautorefname}{Chapter}

\newcommand\spork{\textsc{spork}\xspace}
\newcommand\jdime{\textsc{jdime}\xspace}
\newcommand\automergeptm{\textsc{automergeptm}\xspace}
\newcommand\spoon{\textsc{spoon}\xspace}
\newcommand\linediff{line diff\xspace}
\newcommand\chardiff{character diff\xspace}

\newcommand\casefileheadermerge{\emph{C1}\xspace}
\newcommand\caseconservative{\emph{C2}\xspace}
\newcommand\caseaggressive{\emph{C3}\xspace}
\newcommand\caserename{\emph{C4}\xspace}
\newcommand\caseformatting{\emph{C5}\xspace}

\usepackage{listings}
\def \sporkappendix {\url{https://github.com/slarse/spork-experiments}}
\def \sporkrepo {\url{https://github.com/KTH/spork}}

\newcommand\copyrighttext{%
  \footnotesize \textcopyright 2022 IEEE. Personal use of this material is
  permitted. Permission from IEEE must be obtained for all other uses, in any
  current or future media, including reprinting/republishing this material for
  advertising or promotional purposes, creating new collective works, for
  resale or redistribution to servers or lists, or reuse of any copyrighted
  component of this work in other works.
  DOI: \href{https://doi.org/10.1109/TSE.2022.3143766}{10.1109/TSE.2022.3143766}}
\newcommand\copyrightnotice{%
\begin{tikzpicture}[remember picture,overlay]
\node[anchor=south,yshift=10pt] at (current page.south) {\fbox{\parbox{\dimexpr\textwidth-\fboxsep-\fboxrule\relax}{\copyrighttext}}};
\end{tikzpicture}%
}

\makeatletter
\begin{document}

\title{Spork: Structured Merge for Java with Formatting Preservation}

\author{Simon~Larsén,
        Jean-Rémy Falleri,\\
        Benoit Baudry, ~\IEEEmembership{Member,~IEEE,}
        and~Martin Monperrus,~~\IEEEmembership{Member,~IEEE,}\\~\\
        KTH Royal Institute of Technology, Univ. Bordeaux, Bordeaux INP, CNRS, LaBRI, IUF
        }

\IEEEtitleabstractindextext{\begin{abstract}
    The highly parallel workflows of modern software development have made merging
    of source code a common activity for developers. The state of the practice is
    based on line-based merge, which is ubiquitously used with ``git merge''.
    Line-based merge is however a generalized technique for any text that cannot
    leverage the structured nature of source code, making merge conflicts a common
    occurrence. As a remedy, research has proposed structured merge tools, which
    typically operate on abstract syntax trees instead of raw text. Structured
    merging greatly reduces the prevalence of merge conflicts but suffers from
    important limitations, the main ones being a tendency to alter the formatting
    of the merged code and being prone to excessive running times. In this paper,
    we present \spork, a novel structured merge tool for \textsc{java}.  \spork is
    unique as it preserves formatting to a significantly greater degree than
    comparable state-of-the-art tools. \spork is also overall faster than the state
    of the art, in particular significantly reducing worst-case running times in
    practice. We demonstrate these properties by replaying 1740 real-world file
    merges collected from 119 open-source projects, and further demonstrate several
    key differences between \spork and the state of the art with in-depth case
    studies.
\end{abstract}

}

\markboth{Accepted for publication in IEEE Transactions on Software Engineering}{Shell \MakeLowercase{\textit{et al.}}: Bare Demo of IEEEtran.cls for IEEE Journals}

\maketitle
\copyrightnotice

\IEEEraisesectionheading{\section{Introduction}\label{sec:introduction}}

\IEEEPARstart{B}{ranching} development paths is an unavoidable part of modern
software engineering~\cite{Bird2009}, and developers spend anywhere from a few
hours to several work days each month on integrating changes from
others~\cite{Bird2012}.  This activity is known as ``merging'' code, per the
terminology of mainstream version control systems such as \textsc{git}.  Nearly
all developers use \emph{line-based} merge, which operates on lines of text as
atomic units. It is often referred to as \emph{textual} or \emph{unstructured}
merge~\cite{Mens2002,Le_enich_2014}. This form of merging is simple and
generalizes to any text, but is prone to cause so-called \emph{merge conflicts}
when changes on branches under merge affect the same or adjacent lines. Such
conflicts can be difficult for developers to resolve, and may even cause them
to simply discard a branch that is causing numerous
conflicts~\cite{Nelson2019}.

Merge conflicts are ubiquitous with line-based merge, with conflicts appearing
in about 9\% to 19\% of
merges~\cite{Dias2020,Accioly2017,Brun2013,kasi2013cassandra}. However, many
such conflicts are spurious, because changes on overlapping lines are not
necessarily semantically or syntactically conflicting. This is a fundamental
limitation of line-based merge: it does not capture the underlying structure or
meaning of the text. For example, if two branches add different methods in the
same place in a \textsc{java} file, a line-based merge of said branches yields
a conflict, even though the methods can in fact be safely inserted together in
any order.

To address this spurious conflict problem, the state of the art is
\emph{structured} merge, where the merge process typically acts on
\emph{abstract syntax trees}
(AST)~\cite{Westfechtel1991,Mens2002,Lessenich2012,Le_enich_2014}. Structured
merge has two main advantages: first, it is less influenced by formatting
differences than line-based merge, and second, it can leverage the syntax and
semantics of the considered programming language. For instance, in
\textsc{java}, it is useful for a merge tool to know that duplicated statements
are allowed, but that duplicated fields are not, or that the order of methods
in a class is not important~\cite{Le_enich_2014}.

We observe that the state of the art of structured merge is prone to two main
issues. First, any tool that performs AST transformations must conclude with
\emph{pretty-printing}, which in this context is the act of turning an AST into
its textual representation, i.e. source code. This can fundamentally alter the
formatting of the code~\cite{Waddington2007,Savchenko2019}, which is
undesirable due to the important role that formatting plays in source code
readability~\cite{Buse2008}. Second, structured merge is known for being slow,
with the time complexities of the underlying algorithms often being $O(n^2)$ or
worse~\cite{Le_enich_2014}.

In this paper, we address these issues with a new structured merge tool, called
\spork. \spork is tailored to the \textsc{java} programming language,
leveraging both syntax and semantics of important language constructs to avoid
or resolve conflicts. A key technical novelty of \spork is that it builds upon
the merge algorithm of the \textsc{3dm} merge tool for \textsc{xml}
documents~\cite{Lindholm2004}. In \spork, we both augment the \textsc{3dm}
algorithm, and demonstrate that the core principles are applicable to the
\textsc{java} programming language. As we show in our evaluation, \spork
improves upon the state of the art with respect to the aforementioned problems.
First, \spork reuses source code from the input files when pretty-printing.
This improves formatting preservation over the state of the art in more than
90\% of merged files, with 4 times better preservation in the median case.
Second, \spork's running time performance slightly improves upon the
competition in the median case, but more importantly it significantly reduces
the quantities and magnitudes of the largest running times.

To summarize, our contributions are:

\begin{itemize}
    \item A novel structured merge approach for \textsc{java}, uniquely based on the
        \textsc{3dm} algorithm~\cite{Lindholm2004}, leveraging domain knowledge of
        \textsc{java} to detect and resolve conflicts.
    
    \item \spork, a publicly available prototype implementation: \sporkrepo
    
    \item An evaluation over 890 merge scenarios comprising 1740 file merges, showing that \spork is fast and accurate enough to be used in practice, and formatting preserving. To our knowledge, we are the first to systematically report on formatting preservation for fully AST-based structured merge.

    \item A well-documented benchmarking suite for future research, to study and evaluate \textsc{java} merge tools.

\end{itemize}

\noindent
This article is based on the master's thesis by the first author done at KTH Royal Institute of Technology~\cite{Larsen2020}.

\section{Background}

\subsection{Version Control and Merging}

With the rise in popularity of \emph{distributed version control systems}
(DVCS)~\cite{Brindescu2014,Muslu2014}, the need for merging in software development
has increased~\cite{Bird2009}. The state of the practice is to use unstructured
merge, which operates on raw text, typically using lines of text as atomic
units.  This is fundamentally limited, as a line of text does not represent the
structure of source code. For example, the two lines
\lstinline[breaklines=false]{int a = 2;} and \lstinline[breaklines=false]{int a=2;}
are structurally and semantically equivalent, yet the raw text of those
lines differ by whitespace. The impedance mismatch between lines and code
structure gives rise to needless merge conflicts.

A typical example is if one commit changes the indentation style of some file,
while a parallel one changes the actual code. Such changes are structurally and
semantically compatible, but merging the commits with a line-based merge
results in a merge conflict due to the purely textual differences.

Semistructured merge tools attempt to address some of the problems of
unstructured merge by making use of some structural information in the source
code~\cite{apel2010semistructured,Apel2011,Cavalcanti2015,Cavalcanti2017}. They
work by identifying high-level constructs (e.g. fields and methods) in
the source code, and then merging the content of these modularized units with
line-based merge. Semantic information such as the insignificance of the order
of methods within a type can then be utilized to automatically resolve
conflicts. However, within type members, most notably methods, semistructured
merge still suffers from all of the limitations of unstructured merge.

Structured merge tools go one step further and turn the source code into a
fully resolved
AST~\cite{Lessenich2012,Le_enich_2014,LeBenich2017,Lindholm2004,Zhu2019}. This
allows for a fine-grained merge that respects syntax even within type members,
but also creates a new problem: as the AST abstracts away formatting, the
conversion back from AST to source code, \emph{pretty-printing}, may impose a
completely different formatting style on the merged files. This may be
detrimental to source code quality, as formatting is an integral part of
maintainability and readability~\cite{Buse2008}. In addition, developers care
about the formatting they put in place, as shown by the sheer amount of style
guides that
exist\footnote{\url{https://google.github.io/styleguide/javaguide.html}}\footnote{\url{https://wiki.openjdk.java.net/display/HotSpot/StyleGuide}}\footnote{\url{https://www.cs.cornell.edu/courses/JavaAndDS/JavaStyle.html}}
and the existence of formatting enforcers such as
\textsc{checkstyle}\footnote{\url{https://checkstyle.sourceforge.io/}}.
Current structured merge tools do not preserve formatting, and this poses a
major obstacle for widespread adoption of structured merge. Another problem
with the state of the art in structured merge revolves around running times,
which can become excessive for larger merges due to algorithms with time
complexities that are quadratic or worse~\cite{Le_enich_2014}.

\subsection{The 3DM Algorithm}\label{sec:3dm}

\textsc{3dm} is a state of the art merge algorithm created by
Lindholm~\cite{Lindholm2004}. It performs a \emph{three-way merge} between the
two current revisions (\emph{left} and \emph{right}) of some file, and the
version from which these are derived (the \emph{base})~\cite{Lindholm2004}.
This technique is employed by most merge
tools~\cite{Mens2002,Le_enich_2014,Shen2019}. The most novel part of
\textsc{3dm} is the merge algorithm, which is a generalized merge algorithm for
ordered trees\footnote{The child list of each node is an ordered list of nodes}
that we refer to as \textsc{3dm-merge}. This section presents the theoretical
details of \textsc{3dm-merge} that are relevant for our own work.

\begin{table}[t]
    \centering
    \begin{tabular}{ccc}
        \toprule
        Left      & Base     & Right \\
        \midrule

        add(-a,b,1) & add(a,b) & sum(-a,b,c) \\
    \end{tabular}
    \caption{The left, base and right revisions of a line of code}
    \label{tab:3dm:function_call_revisions}
\end{table}

\begin{figure}[t]
    \begin{center}
        \begin{tikzpicture}[sibling distance=4.05em, every node/.style={rectangle, draw=black, align=center, thick}, every label/.style={rectangle,draw=none}]
            \small

            \begin{scope}
                \tikzstyle{edge from parent}=[draw,black,thick,->]
                \node[label={[label distance=0.25cm, fill=lightgray]north:{Base ($T_0$)}}] (base_call) {call$_{01}$(add)}
                child {node (base_lhs) {ref$_{02}$(a)}}
                child {node (base_rhs) {ref$_{03}$(b)}};

                \node[label={[label distance=0.25cm, fill=lightgray]north:{Left ($T_1$)}},below left=0.5cm and 0.6cm of base_call] (left_call) {call$_{11}$(add)}
                child {node (left_uop) {uop$_{12}$(-)}
                    child {node (left_lhs) {ref$_{13}$(a)}}
                }
                child {node (left_mid) {ref$_{14}$(b)}}
                child {node (left_rhs) {int$_{15}$(1)}};

                \node[label={[label distance=0.25cm, fill=lightgray]north:{Right} ($T_2$)},below right=0.5cm and 0.6cm of base_call] (right_call) {call$_{21}$(sum)}
                child {node (right_uop) {uop$_{22}$(-)}
                    child {node (right_lhs) {ref$_{23}$(a)}}
                }
                child {node (right_mid) {ref$_{24}$(b)}}
                child {node (right_rhs) {ref$_{25}$(c)}}
                ;
            \end{scope}

            \path [thick, dotted, blue, <->]
            (base_call) edge (left_call)
            (base_lhs) edge [bend left=95] (left_lhs)
            (base_rhs.south) edge (left_mid.north)
            ;

            \path [thick, densely dashed, red, <->]
            (base_call) edge (right_call)
            (base_lhs.south) edge (right_lhs)
            (base_rhs) edge (right_mid)
            ;

            \path [thick,dash pattern=on 1pt off 2pt on 2pt off 2pt,olive,<->]
            (left_call) edge [bend left=10] (right_call)
            (left_lhs) edge [bend right=15] (right_lhs)
            (left_uop) edge [bend right=30] (right_uop)
            (left_mid.south) edge [bend right=30] (right_mid.south)
            ;
        \end{tikzpicture}

        \begin{tikzpicture}
            \begin{axis}[hide axis,
                    width=5cm,
                    height=2cm,
                    xmin=1,
                    xmax=1,
                    ymin=1,
                    ymax=1,
                    legend style={nodes={rectangle,draw=none},draw=none,legend cell align=left}
                ]
                \addlegendimage{black,thick,->}
                \addlegendentry{Parent/child relationship};
                \addlegendimage{red,thick,densely dashed,<->}
                \addlegendentry{Base/right matching};
                \addlegendimage{blue,thick,dotted,<->}
                \addlegendentry{Base/left matching};
                \addlegendimage{olive,thick,dash pattern=on 1pt off 2pt on 2pt off 2pt,<->}
                \addlegendentry{Left/right matching};
            \end{axis}
        \end{tikzpicture}
    \end{center}
    \caption{Pairwise matchings between simplified ASTs of the function calls in \autoref{tab:3dm:function_call_revisions}.
    Each node is subscripted with a unique identifier $xy$, where $x$ indicates
the tree the node belongs to and $y$ is unique within tree $x$.}
    \label{fig:three_way_tree_matching}
\end{figure}

\begin{table}[t]
    \setlength\tabcolsep{4pt}
    \caption{PCS set of $T_0$ in \autoref{fig:three_way_tree_matching}, ordered into
    child lists}
    \label{tab:pcs_child_lists}
    \begin{center}
        \begin{tabular}{ll}
            \toprule
            Node & PCS child list\\
            \midrule
            $\bot$ & $(\bot, \dashv, call_{01}),\ (\bot, call_{01}, \vdash)$ \\
            $call_{01}$ & $(call_{01}, \dashv, ref_{02}),\ (call_{01}, ref_{02}, ref_{03}),\ (call_{01}, ref_{03}, \vdash)$ \\
            $ref_{02}$ & $(ref_{02}, \dashv, \vdash)$ \\
            $ref_{03}$ & $(ref_{03}, \dashv, \vdash)$
        \end{tabular}
    \end{center}
\end{table}

\begin{table}[t]
    \centering
    \setlength\tabcolsep{4pt}
    \footnotesize
    \caption{Class representatives mapping for \autoref{fig:three_way_tree_matching}}
    \label{tab:class_rep_map}
    \begin{tabular}{lccccccccccccc}
        Node ID      & 01 & 02 & 03 & 11 & 12 & 13 & 14 & 15 & 21 & 22 & 23 & 24 & 25 \\
        \midrule                                                                 
        Class rep.   & 01 & 02 & 03 & 01 & 12 & 02 & 03 & 15 & 01 & 12 & 02 & 03 & 25 \\
    \end{tabular}
\end{table}

\subsubsection{Data Structures of 3DM}\label{sec:change_set}
\textsc{3dm-merge} does not operate on a traditional tree structure, but on an
abstract representation of an ordered tree, called a \emph{change
set}~\cite{Lindholm2004}. This is composed of two primary data types. The first
of these is the \emph{parent-child-successor} (PCS) triple, which encodes the
structure of the tree. A PCS triple is written in the form
$pcs(parent,pred,succ)$, where $parent$ is an arbitrary tree node,
$pred$\footnote{In the original paper, this node is called $child$} is any of
$parent$'s children, and $succ$ is the node in $parent$'s child list that
directly succeeds $pred$. For a given tree, the set of triples with $parent=x$
therefore encode the child list of $x$.

There are also three kinds of \emph{virtual} nodes: a virtual root $\bot$, a
virtual start of a child list $\dashv$ and a virtual end of a child list
$\vdash$~\cite{Lindholm2004}. These nodes mark the boundaries of the tree's
structure. As an example of applying the PCS concepts, consider the base
revision function call in \autoref{tab:3dm:function_call_revisions} and its
corresponding AST $T_0$ in \autoref{fig:three_way_tree_matching}, and how the
syntactical structure is fully encoded by the PCS set in
\autoref{tab:pcs_child_lists}. Note that nodes are identified by ID, and not by
content. For example, a variable reference $ref_{01}(a)$ is different from
another reference $ref_{02}(a)$, even though they are identical apart from ID.
We often refer to nodes by their IDs alone to reduce the verbosity of figures
and tables. For example, the PCS set in \autoref{tab:pcs_child_lists} is
equivalent to the base revision PCS set in \autoref{tab:3dm:pcs_sets}.

The second data type is the \emph{content} tuple, written $c(v,m)$, where $v$
is any concrete node and $m$ is $v$'s content, the exact form of which is
domain dependent. In general, the content $m$ of a tree node $v$ is all data
related to $v$ that does not impact the structure of the tree. We express $m$
as a set of values. As a concrete example, the content tuples of $T_0$ in
\autoref{fig:three_way_tree_matching} is $\{c(01, "add"), c(02, "a"), c(03,
"b")\}$. Note that $m$ is a set. If for example the base and left revision of
\lstinline{add} had modifiers \lstinline{public} and \lstinline{private},
respectively, then the content tuples would be $c(01, \{public, "add"\})$ and
$c(11, \{private, "add"\})$. The change set is simply the union of the content
tuples and PCS triples of a tree.

A change set is said to be \emph{consistent} if each node $v$ has at most 1) one
parent $x$, 2) one predecessor $y$, 3) one successor $z$ and 4) one content set
$m$~\cite{Lindholm2004}. A consistent change set is unambiguous, and a tree always
encodes a consistent change set. As the consistency criteria allow a node to
have less than one parent, predecessor, successor and content set, a consistent
change set does not necessarily encode a well-formed tree.

\subsubsection{Matchings and Class Representatives}\label{sec:3dm:class_reps}
\textsc{3dm-merge} makes use of tree matchings to determine where trees to be
merged are similar~\cite{Lindholm2004}. We define a tree matching as a symmetric
relation between the nodes of two trees $T_i$ and $T_j$, where each node $v \in
T_i$ can be matched to at most one node $w \in T_j$.  The details of how a match
between nodes $v$ and $w$ is computed varies greatly between matching
algorithms. The most powerful algorithms consider many factors, including the
nodes' relative positions, their contents, as well as the similarities of their
subtrees~\cite{Falleri2014}.

For a three-way merge, three pairwise matchings are typically required:
base/left, base/right, and left/right. \autoref{fig:three_way_tree_matching}
illustrates this for a simple merge scenario. Note for example that the
base/left matching contains a node matching between root nodes $01$ and $11$,
which is reasonable as the root nodes of the base and left revisions are
identical and in the same position. Note also that the base/right matching
contains a node matching between root nodes $01$ and $21$ even though the method
names differ, as their positions and subtrees are similar enough.

The tree matchings are then used to create a \emph{class representatives}
mapping~\cite{Lindholm2004}. Each node $v$ is mapped to precisely one class
representative $w$, which we denote with $(v \rightarrow w)$ and refer to
as a \emph{classmapping}. All nodes assigned to the same class representative
are considered equivalent by \textsc{3dm-merge}. Formally, let $T_0$, $T_1$ and
$T_2$ be the base, left and right revisions, respectively. A node $v \in T_i$ is
classmapped to $w \in T_j$, i.e. $(v \rightarrow w)$, if the following
three criteria are met: 1) $v$ is matched to $w$, 2) $j \leq i$ and 3) there is
no other node $u \in T_k$ where $v$ is matched to $u$ and $k < j$. Note that a
classmapping is directional, so $(v \rightarrow w) \centernot\implies (w
\rightarrow v)$.

Intuitively, a node without any matches in other trees is classmapped to itself,
and the base revision takes precedence over the left revision, which in turn
takes precedence over the right. This is evident from
\autoref{tab:class_rep_map}, which shows the class representatives mapping
produced from the matchings in \autoref{fig:three_way_tree_matching}. For
example, all nodes in the base revision are classmapped to themselves, and $(21
\rightarrow 01)$ even though there is also a node matching between $11$
and $21$, showing the base revision's precedence. Similarly, $(22 \rightarrow
12)$ instead of the other way around, showing the left revisions
precedence over the right.

\begin{table}[t]
    \setlength\tabcolsep{2pt}
    \centering
    \caption{PCS sets of the trees in
        \autoref{fig:three_way_tree_matching}, the raw merge of these and
        the finished merge. All nodes are presented as their class
        representative IDs.}
    \label{tab:3dm:pcs_sets}
    \begin{tabular}{ll}
        \toprule
        Revision        & PCS set \\
        \midrule
        Left        & $(\bot,\dashv,01),(\bot,01,\vdash),$ \\
                    & $(01,\dashv,12),(01,12,03),(01,03,15),(01,15,\vdash),$ \\
                    & $(12,\dashv,02),(12,02,\vdash),(02,\dashv,\vdash),(03,\dashv,\vdash),(15,\dashv,\vdash)$ \\
        \midrule
        Base        & $(\bot,\dashv,01),(\bot,01,\vdash),$ \\
                    & $(01,\dashv,02),(01,02,03),(01,03,\vdash),(02,\dashv,\vdash),(03,\dashv,\vdash)$ \\
        \midrule
        Right       & $(\bot,\dashv,01),(\bot,01,\vdash),$ \\
                    & $(01,\dashv,12),(01,12,03),(01,03,25),(01,25,\vdash),$ \\
                    & $(12,\dashv,02),(12,02,\vdash),(02,\dashv,\vdash),(03,\dashv,\vdash),(25,\dashv,\vdash)$ \\
        \midrule
        Raw merge   & $(\bot,\dashv,01),(\bot,01,\vdash),$ \\
                    & $(01,\dashv,02),(01,02,03),(01,03,\vdash),(01,\dashv,12),$ \\
                    & $(01,12,03),(01,03,15),(01,15,\vdash),$ \\
                    & $(01,03,25),(01,25,\vdash),(12,\dashv,02),(12,02,\vdash),$ \\
                    & $(02,\dashv,\vdash),(03,\dashv,\vdash),(15,\dashv,\vdash),(25,\dashv,\vdash)$ \\
        \midrule
        Merge       & $(\bot,\dashv,01),(\bot,01,\vdash),$ \\
                    & $(01,\dashv, 12),(01,12,03),(01,03,15),(01,15,\vdash),$ \\
                    & $(01,03,25),(01,25,\vdash),(12,\dashv,02),(12,02,\vdash),$ \\
                    & $(02,\dashv,\vdash),(03,\dashv,\vdash),(15,\dashv,\vdash),(25,\dashv,\vdash)$ \\
    \end{tabular}
\end{table}

\subsubsection{Merging in 3DM}\label{sec:3dm:merge}
\textsc{3dm-merge} operates in two distinct phases. First, it converts the AST
revisions into change sets, with each node mapped to its class representative,
and initializes the \emph{raw merge} as the set union of these change sets.
Unless all input revisions are identical, the raw merge always contains
violations of the consistency criteria presented in \autoref{sec:change_set},
so-called \emph{inconsistencies}. For example, the two PCS triples $pcs(x,y,z)$
and $pcs(x',y,z)$ violate the criterion that each node has a unique parent. The
second and most important phase of \textsc{3dm-merge} is dedicated to finding
and removing inconsistencies, with the end-goal of turning the raw merge into a
consistent change set.

Consider \autoref{tab:3dm:pcs_sets}, which shows an example merge of the
trees in \autoref{fig:three_way_tree_matching} in terms of PCS elements only.
Note how the identical elements of the revisions are merged simply by the
nature of a set union, such as $(01,\dashv,12)$ that is present in both
left and right revisions, yet only appears once in the raw merge. The raw merge
also contains numerous inconsistencies that need to be processed.
\textsc{3dm-merge} identifies these by iterating over each element $\delta$ of
the change set, and searching for another element $\delta'$ such that $\delta$
and $\delta'$ are inconsistent. For example, given
$\delta=(01,\dashv,02)$, then $\delta'=(01,\dashv,12)$ is found to be
inconsistent. Note that $\delta$ is present in the base revision, while
$\delta'$ is not. The inconsistency can therefore be resolved by removing
$\delta$, thus preserving the change represented by $\delta'$. We refer to such
an inconsistency as \emph{soft}.

However, now consider the inconsistent pair $\delta=(01,03,15)$ and
$\delta'=(01,03,25)$. Neither element is present in the base revision, and
therefore removing either would cause change information to be lost. We refer to
this as a \emph{hard} inconsistency, and these are always caused by a conflict
on the AST level. In this case, the conflict is caused by the left and right
revisions inserting the nodes $int_{15}(1)$ and $ref_{25}(c)$ in the same place.
Note the terminology used; \emph{conflict} refers to incompatible changes to the
ASTs, and \emph{inconsistency} refers to a violation of the consistency criteria
in the change set.

The same principles apply to content inconsistencies. For example, there is a
content inconsistency between $c(01,"add")$ and $c(01,"sum")$\footnote{Note
that nodes are mapped to class representatives}. This is a soft inconsistency
as $c(01,"add")$ is present in the base revision, which can therefore be
removed. Hard content inconsistencies are analogous to hard PCS
inconsistencies, and occur when neither of the inconsistent elements are
present in the base revision.

\begin{figure*}
    \centering
    \begin{tikzpicture}[
            file/.style={rectangle, draw=gray, align=center, minimum width=30, thin},
            ast/.style={rectangle, draw=gray, align=center, minimum width=30, thin},
            matching/.style={rectangle, draw=gray, align=center, minimum width=60, thin},
            phase/.style={rectangle, draw=black, align=center, thick},
            outline/.style={ellipse, draw=black, align=center, thin, dotted, minimum width=55, minimum height=90},
            library/.style={ellipse, draw=gray, thick, align=center},
            branch/.style={diamond, draw=black, thin, align=center}
        ]
        \node[file] (f_left) {$f_{left}$};
        \node[file, below=0.1cm of f_left] (f_base) {$f_{base}$};
        \node[file, below=0.1cm of f_base] (f_right) {$f_{right}$};

        \node[phase, right=2cm of f_base] (parse) {Parse};

        \node[ast, right=3cm of parse] (t_base) {$T_{base}$};
        \node[ast, above=0.1cm of t_base] (t_left) {$T_{left}$};
        \node[ast, below=0.1cm of t_base] (t_right) {$T_{right}$};

        \node[phase, right=2cm of t_base] (match) {Match};

        \node[matching, right=2cm of match] (m_leftright) {$M_{left \Leftrightarrow right}$};
        \node[matching, above=0.1cm of m_leftright] (m_baseleft) {$M_{base \Leftrightarrow left}$};
        \node[matching, below=0.1cm of m_leftright] (m_baseright) {$M_{base \Leftrightarrow right}$};

        \node[outline, right=2.58cm of parse, label={[label distance=0.25cm, fill=lightgray]north:{ASTs}}] (trees) {};
        \node[outline, left=1.58cm of parse, label={[label distance=0.25cm, fill=lightgray]north:{Files (input)}}] (files) {};
        \node[outline, minimum width=100, right=1.35cm of match, label={[label distance=0.25cm, fill=lightgray]north:{Matchings}}] (matchings) {};

        \node[phase, below=1cm of matchings] (merge) {\textsc{spork-3dm}};
        \node[ast, left=0.5cm of merge] (changeset) {$ChangeSet$};

        \node[phase, below left=-1.5cm and 0.5cm of changeset] (buildast_large) {
            \centering
            \begin{minipage}[b][2.7cm]{9cm}
                Handle conflicts and build AST
            \end{minipage}
        };

        \node[phase, below left=0.5cm and 1cm of changeset] (buildast) {Build \textsc{AST}};
        \node[ast, left=0.4cm of buildast] (t_merge) {$T_{merge}$};

        \node[branch, above left=0.25cm and 3.5cm of t_merge] (duplicates) {DTM?};
        
        \node[ast, right=1.5cm of duplicates] (type_member_left) {$T_{L\_dup}$};
        \node[ast, right=0.1cm of type_member_left] (type_member_base) {$T_{empty}$};
        \node[ast, right=0.1cm of type_member_base] (type_member_right) {$T_{R\_dup}$};
        \node[outline, minimum width=150, minimum height=30, right=-3.2cm of type_member_base, label={[label distance=0.1cm, fill=lightgray]north:{Duplicated members}}] (duplicated_members) {};

        \node[phase, above left=-1cm and 1.5cm of buildast_large] (print) {Pretty-print};
        \node[file, below=.5cm of print, label={[label distance=0.25cm, fill=lightgray]south:{Output}}] (f_merge) {$f_{merge}$};

        \node[library, below=0.5cm of parse] (spoon) {\textsc{spoon}};
        \node[library, above=1cm of match] (gumtree) {\textsc{gumtree}};

        \path [black, thick, dotted, ->]
        (spoon.west) edge (print)
        (spoon) edge (parse)
        (spoon.east) edge (buildast_large)
        (gumtree) edge (match)
        ;

        \path [thick, gray, ->]
        (files) edge (parse)
        (parse) edge (trees)
        (trees) edge (match)
        (match) edge (matchings)

        (trees) edge (merge.north)
        (matchings) edge (merge)
        (merge) edge (changeset)
        (changeset) edge (buildast.east)
        (buildast) edge (t_merge)
        (t_merge.west) edge (duplicates.south)
        (duplicates.east) edge [below] node {yes} (duplicated_members.west)
        (duplicated_members.east) edge (match.south)
        (duplicated_members.east) edge [bend left=20] (merge)
        (duplicates.west) edge [above] node {no} (print)
        (print) edge (f_merge)
        ;
    \end{tikzpicture}

    \begin{tikzpicture}
        \begin{axis}[hide axis,
                width=5cm,
                height=2cm,
                xmin=1,
                xmax=1,
                ymin=1,
                ymax=1,
                legend style={nodes={rectangle,draw=none},draw=none,legend cell align=left}
            ]
            \addlegendimage{gray,thick,->}
            \addlegendentry{Data flow};
            \addlegendimage{black,thick,dotted,->}
            \addlegendentry{Library used in};
        \end{axis}
    \end{tikzpicture}

    \caption{Schematic drawing of \spork's phases. Thin-lined
    rectangles represent data, filled rectangles are labels describing the
    closest data, and related data is grouped within dotted outlines. Thick-lined
    rectangles represent phases and ellipses represent libraries. 
    ``DTM''=duplicated type member.}
    \label{fig:spork:architecture}
\end{figure*}

\section{Technical contribution: Spork}\label{sec:spork}

\spork performs structured merging of \textsc{java} files in 5 distinct phases,
which are illustrated schematically in \autoref{fig:spork:architecture}. The first
phase consists of parsing source files into ASTs, as described in
\autoref{sec:spork:parsing}. It is followed by a matching phase, in which tree
matchings are computed as described in \autoref{sec:spork:tree_matching}.
These matchings are used in \spork's variation of \textsc{3dm-merge}, as
described in \autoref{sec:spork:merge}. \spork then enters a composite phase in
which it handles conflicts and builds a merged AST, as described in
\autoref{sec:spork:conflicts}. The final phase is responsible for
pretty-printing the merged AST, as described in
\autoref{sec:spork:pretty_printing}.

\subsection{Parsing}
\label{sec:spork:parsing}
\spork uses the \textsc{spoon} library~\cite{Pawlak2015} to parse \textsc{java}
source files into ASTs.
In our example, Spoon is responsible for going from the raw source code in
\autoref{tab:3dm:function_call_revisions} to their corresponding ASTs in
\autoref{fig:three_way_tree_matching}
At this point, \spork identifies and stores the style of
indentation in the source file as the amount of tabs or spaces that precede
top-level type members. This is necessary to later be able to print the merged
file with the correct indentation. \textsc{spoon} itself also stores the
original source code of each parsed file, which \spork in certain cases
directly reuses to respect arbitrary formatting styles. This is further
detailed in \autoref{sec:spork:pretty_printing}.

\subsection{Tree Matching}\label{sec:spork:tree_matching}

\spork uses
\textsc{gumtree}\footnote{\url{https://github.com/gumtreediff/gumtree}}$^,$\footnote{ \url{https://github.com/spoonlabs/gumtree-spoon-ast-diff}}~\cite{Falleri2014}
to compute the pairwise base/left, base/right and left/right matchings between
the ASTs. \textsc{gumtree} for example produces the matchings shown in
\autoref{fig:three_way_tree_matching}.

The base/left and base/right matchings allow \spork to align the two changed
revisions with the base. The left/right matching is primarily used to
merge identical or near-identical additions in the left and right revisions.
These matchings ground the set properties utilized in the raw merge, as shown
in~\autoref{tab:3dm:pcs_sets}.

\subsection{Merging Approach}\label{sec:spork:merge}
Merging is the primary technical contribution in \spork, as most of the
functionality is implemented directly in the tool itself. The implementation is
based on \textsc{3dm-merge}, which is described in \autoref{sec:3dm}.

\subsubsection{Mapping to class representatives}\label{sec:spork:class_representatives}
The mapping to class representatives largely follows the theoretical ideas
presented in \autoref{sec:3dm:class_reps}. First, nodes are classmapped to
themselves if they are unmatched, or to at most one node in another revision
according to the matching prioritization presented in
\autoref{sec:3dm:class_reps}.
Ultimately, this results in a class representatives mapping like that of
\autoref{tab:class_rep_map}.

The left/right matching is however prone to contain spurious matchings, and
carelessly adding these to the class representatives mapping can cause
unnecessary conflicts. For example, if the left and right revisions add
identical method parameters to different methods, the parameters may be matched
due to their similarity, even though they are entirely unrelated. To reduce the
effect of spurious matchings, \spork implements two heuristics to decide
whether or not to classmap $(v_{right} \rightarrow v_{left})$ given a match
between $v_{right}$ in the right revision and $v_{left}$ in the left revision.
First, the matching is ignored if any of the nodes are already classmapped to a
node $v_{base}$ in the base revision. This prevents the classmappings
$(v_{left} \rightarrow v_{base})$ and $(v_{right} \rightarrow v_{left})$ from
coexisting, which avoids a strange situation in which a node from the left
revision appears in the right revision's change set, but not the left
revision's.  Second, the parents of $v_{left}$ and $v_{right}$ must already be
classmapped to the same class representative. This prevents unrelated matchings,
such as matchings between method parameters of different methods, from making
their way into the class representatives mapping.

Adding eligible left/right matches to the class representatives is
performed with a top-down scan of the left tree. The fact that the scan is
top-down is important, as it allows arbitrarily complex subtrees to be
incrementally mapped as long as their roots have parents that are already
mapped to the same class representative.

\subsubsection{Converting an AST to a change set}\label{sec:spoon_to_pcs} A
tree can be converted into a change set by traversing it top-down and creating
a PCS child list for each node, as well as extracting each node's content. This
corresponds to the process of going from the base tree in
\autoref{fig:three_way_tree_matching} to the PCS triples in
\autoref{tab:pcs_child_lists} and the associated content set shown in
\autoref{sec:change_set}.

For some complex nodes, naively building a PCS child list of their
direct children is insufficient to achieve appropriate separation between
distinct syntactical elements. For example, consider the method declaration in
\autoref{lst:spork:java_meth_decl}. In \textsc{spoon}, parameters and
thrown types are considered direct children of the method node, and so appear
in its child list. \autoref{fig:spork:naive_child_list} shows a schematic AST
with a naively built child list for the method node in
\autoref{lst:spork:java_meth_decl}. As the end of the list of parameters is
adjacent to the beginning of the list of thrown types, structural modifications
to the former may conflict with structural modifications to the latter.

\begin{lstlisting}[
basicstyle=\footnotesize,
language=java,
numbers=none,
label={lst:spork:java_meth_decl},
caption={A \textsc{java} method declaration},
xleftmargin=0cm,float
]
int div(int lhs, int rhs) throws ArithmeticException { ... }
\end{lstlisting}

To avoid collections of elements of different types within a child list to
conflict with each other, \spork inserts \emph{intermediate virtual
nodes} when building the PCS structure. A schematic example of this is shown in
\autoref{fig:spork:virtual_node_child_list}, where the \texttt{parameters} and
\texttt{thrown} virtual nodes separate the previously adjacent parameters and
thrown types. It is important that all applicable intermediate virtual nodes
are inserted even if the parent node has no children of the corresponding
types, as otherwise conflicts can occur due to insertions and deletions of the
virtual nodes themselves.

\begin{figure}[t]
    \small
    \begin{center}
        \begin{tikzpicture}[every node/.style={rectangle, draw=black, thick}]
            \node (name) {\lstinline{div()}};
            \node[below left=0.7cm and 3.5cm of name] (ret) {\lstinline{int}};
            \node[right=0.3cm of ret] (param1) {\lstinline{int lhs}};
            \node[right=0.3cm of param1] (param2) {\lstinline{int rhs}};
            \node[right=0.3cm of param2] (thrown) {\lstinline{ArithmeticException}};
            \node[right=0.3cm of thrown] (body) {\lstinline{()}};
            \node[below=0.5cm of body] (statements) {\lstinline{...}};

            \path [thick, black, ->]
            (name) edge (param1.north)
            (name) edge (param2.north)
            (name) edge (ret.north)
            (name) edge (thrown)
            (name) edge (body.north)
            (body) edge (statements)
            ;
        \end{tikzpicture}
    \end{center}
    \caption{Schematic drawing of naively built child list for the method
    declaration in \autoref{lst:spork:java_meth_decl}}
    \label{fig:spork:naive_child_list}
\end{figure}

\begin{figure}[t]
    \small
    \begin{center}
        \begin{tikzpicture}[every node/.style={rectangle, draw=black, thick}]
            \node (name) {\lstinline{div()}};
            \node[below left=0.7cm and 2.3cm of name] (ret) {\lstinline{int}};
            \node[right=0.3cm of ret] (params) {\lstinline{parameters}};
            \node[below left=0.7cm and 0.3cm of params] (param1) {\lstinline{int lhs}};
            \node[right=0.3cm of param1] (param2) {\lstinline{int rhs}};
            \node[right=0.3cm of params] (thrown_types) {\lstinline{thrown}};
            \node[right=0.3cm of param2] (thrown) {\lstinline{ArithmeticException}};
            \node[right=0.3cm of thrown_types] (body) {\lstinline{()}};
            \node[below=0.3cm of body] (statements) {\lstinline{...}};

            \path [thick, black, ->]
            (name) edge (params)
            (params) edge (param1)
            (params) edge (param2)
            (name) edge (ret.north)
            (name) edge (thrown_types)
            (thrown_types) edge (thrown)
            (name) edge (body.north)
            (body) edge (statements)
            ;
        \end{tikzpicture}
    \end{center}
    \caption{Schematic drawing of the method declaration in
    \autoref{lst:spork:java_meth_decl} with intermediate virtual nodes for the parameters and thrown types}
    \label{fig:spork:virtual_node_child_list}
\end{figure}

\subsubsection{Core \spork Algorithm}
\label{sec:spork:3dm_merge}

\begin{figure}[t]
    \removelatexerror

    \begin{algorithm}[H]
\SetKwProg{Fn}{Function}{ is}{end}
\DontPrintSemicolon
\caption{Spork-3dm}
\SetKw{PyGt}{$>$}\SetKw{PyIn}{$\in$}\SetKw{PyIsnot}{$\ne$}\SetKw{PyIs}{is}\SetKw{PyLte}{$\leq$}\SetKw{PyNotin}{$\notin$}
\SetKwFunction{copyOf}{copyOf}\SetKwFunction{getAllInconsistentPcs}{getAllInconsistentPcs}\SetKwFunction{getContentTuples}{getContentTuples}\SetKwFunction{handleContent}{handleContent}\SetKwFunction{hardContentInconsistency}{hardContentInconsistency}\SetKwFunction{hardPcsInconsistency}{hardPcsInconsistency}\SetKwFunction{merge}{merge}\SetKwFunction{removePcs}{removePcs}\SetKwFunction{removeSoftContentInconsistencies}{removeSoftContentInconsistencies}\SetKwFunction{removeSoftPcsInconsistencies}{removeSoftPcsInconsistencies}\SetKwFunction{setContentTuples}{setContentTuples}\SetKwFunction{size}{size}\SetKwFunction{toChangeSet}{toChangeSet}\SetKwFunction{unionOf}{unionOf}
\SetKwData{baseCS}{baseCS}\SetKwData{base}{base}\SetKwData{classReps}{classReps}\SetKwData{cts}{cts}\SetKwData{ct}{ct}\SetKwData{inconsistencies}{inconsistencies}\SetKwData{leftCS}{leftCS}\SetKwData{left}{left}\SetKwData{mergeCS}{mergeCS}\SetKwData{nonBaseCts}{nonBaseCts}\SetKwData{otherPcs}{otherPcs}\SetKwData{parent}{parent}\SetKwData{pcsSet}{pcsSet}\SetKwData{pcs}{pcs}\SetKwData{pred}{pred}\SetKwData{rightCS}{rightCS}\SetKwData{right}{right}\SetKwData{succ}{succ}\SetKwData{tree}{tree}
\Fn{\merge} {
\KwData{base, left, right: Tree, classReps: Map[Tree, Tree]}
\KwResult{ChangeSet}
  \baseCS = \toChangeSet{\base, \classReps}\;
  \leftCS = \toChangeSet{\left, \classReps}\;
  \rightCS = \toChangeSet{\right, \classReps}\;
  \mergeCS = \unionOf{\baseCS, \leftCS, \rightCS}\;
  \For { \pcs \PyIn \copyOf{\mergeCS.\pcsSet} }
  {
    \removeSoftPcsInconsistencies{\pcs, \mergeCS, \baseCS}\;
    \handleContent{\pcs, \mergeCS, \baseCS}\;
  }
  \Return\mergeCS\;
}
\Fn{\removeSoftPcsInconsistencies} {
\KwData{pcs: PCS, mergeCS, baseCS: ChangeSet}
  \inconsistencies = \getAllInconsistentPcs{\pcs, \mergeCS}\;
  \uIf {\size{\inconsistencies} \PyIs 0}
  {
    \Return\;
  }
  \uIf {\pcs \PyIn \baseCS}
  {
    \removePcs{\mergeCS, \pcs}\;
    \Return\;
  }
  \For { \otherPcs \PyIn \inconsistencies }
  {
    \uIf {\otherPcs \PyIn \baseCS}
    {
      \removePcs{\mergeCS, \otherPcs}\;
    }
    \uElse
    {
      \hardPcsInconsistency{\pcs, \otherPcs}\;
    }
  }
}
\Fn{\handleContent} {
\KwData{pcs: PCS, mergeCS, baseCS: ChangeSet}
  \For { \tree \PyIn \{\pcs.\parent, \pcs.\pred, \pcs.\succ\} }
  {
    \removeSoftContentInconsistencies{\tree, \mergeCS, \baseCS}\;
  }
}
\Fn{\removeSoftContentInconsistencies} {
\KwData{tree: Tree, mergeCS, baseCS: ChangeSet}
  \cts = \getContentTuples{\tree, \mergeCS}\;
  \uIf {\size{\cts} \PyLte 1}
  {
    \Return\;
  }
  \nonBaseCts = \{\ct $|$ \ct $\in$ \cts, \ct \PyNotin \baseCS\}\;
  \setContentTuples{\tree, \nonBaseCts, \mergeCS}\;
  \uIf {\size{\nonBaseCts} \PyGt 1}
  {
    \hardContentInconsistency{\nonBaseCts}\;
  }
}
\end{algorithm}
 
    \caption{Pseudocode for \textsc{spork-3dm}}\label{fig:spork:merge}
\end{figure}

\begin{figure}[t]
    \removelatexerror

    \begin{algorithm}[H]
\SetKwProg{Fn}{Function}{ is}{end}
\DontPrintSemicolon
\caption{Helpers for spork-3dm}
\SetKw{PyGt}{$>$}\SetKw{PyIn}{$\in$}\SetKw{PyIsnot}{$\ne$}\SetKw{PyIs}{is}\SetKw{PyLte}{$\leq$}\SetKw{PyNotin}{$\notin$}
\SetKwFunction{getAllInconsistentPcs}{getAllInconsistentPcs}\SetKwFunction{getContentTuples}{getContentTuples}\SetKwFunction{hardContentInconsistency}{hardContentInconsistency}\SetKwFunction{hardPcsInconsistency}{hardPcsInconsistency}\SetKwFunction{setContentTuples}{setContentTuples}\SetKwFunction{toChangeSet}{toChangeSet}

\Fn{\toChangeSet} {
\KwData{tree: Tree, classReps: Map[Tree, Tree]}
\KwResult{ChangeSet}
  Convert tree to a change set with nodes mapped to their class representatives.
}
\Fn{\getAllInconsistentPcs} {
\KwData{pcs: PCS, cs: ChangeSet}
\KwResult{List[PCS]}
  Get all parent, predecessor and successor inconsistencies for pcs in cs.
}
\Fn{\getContentTuples} {
\KwData{tree: Tree, changeSet: ChangeSet}
\KwResult{Set[ContentTuple]}
  Get all content tuples related to the tree according to the change set. 
}
\Fn{\setContentTuples} {
\KwData{tree: Tree, contents: Set[ContentTuple], changeSet: ChangeSet}
  Set the content tuples associated with the tree in the change set. 
}
\Fn{\hardPcsInconsistency} {
\KwData{pcs: PCS, other: PCS}
  Register pcs and other as a hard inconsistency. 
}
\Fn{\hardContentInconsistency} {
\KwData{contentTuples: Set[ContentTuple]}
  Register the provided content tuples as a hard inconsistency. 
}
\end{algorithm}
 
    \caption{Helper function definitions for \textsc{spork-3dm}}\label{fig:spork:helpers}
\end{figure}

\spork implements a non-trivial variation of \textsc{3dm-merge}, called \textsc{spork-3dm}.
The key concepts of the algorithm are presented with pseudocode in
\autoref{fig:spork:merge}. To ease understanding, non-obvious helper functions
are described in \autoref{fig:spork:helpers}.

The \lstinline{merge} function of \autoref{fig:spork:merge} shows the
\textsc{spork-3dm} algorithm at a high level of abstraction. It starts out with
converting the input trees to change sets, all nodes being mapped to their
class representatives. The union of these change sets forms the initial raw
merge \lstinline{mergeCS}, which as noted in \autoref{sec:3dm:merge} may
contain inconsistencies for any non-trivial merge. The loop starting on line 6
is concerned with making \lstinline{mergeCS} consistent by removing soft
inconsistencies, and recording any hard inconsistencies. Acting on hard
inconsistencies, known as \emph{conflict handling}, is not part of
\textsc{3dm-merge} nor \textsc{spork-3dm} and \spork's conflict handling is
described in \autoref{sec:spork:conflicts}.

The \lstinline{removeSoftPcsInconsistencies} function is the heart of the
algorithm: if two PCS are found to be inconsistent, \spork removes any that is in the
base revision. If neither is in the base revision, they are in a hard
inconsistency. Note that each invocation of the function is concerned only with
PCS that are inconsistent with the input variable \lstinline{pcs}.  Therefore,
if the input \lstinline{pcs} is found to be in the base revision, all
inconsistencies related to it are resolved by removing it from the change set,
hence the early return on line 16. \autoref{tab:spork:dry_run} shows the
effects of a series of invocations of this function on the raw merge of
\autoref{tab:3dm:pcs_sets}, until the final merge is achieved.

The \lstinline{removeSoftContentInconsistencies} function follows the
same principles as \lstinline{removeSoftPcsInconsistencies}. It is however
simplified due to each content tuple belonging to precisely one tree node, and
each tree node having at most one content tuple from each revision. A hard
content inconsistency is therefore local to the node in which it occurs, and is
identified by there being more than one non-base content tuple. In a
three-way merge, the only possibility is that there are precisely two non-base
content tuples: one from the left revision, and one from the right.

\textsc{spork-3dm} differs from \textsc{3dm-merge} in two key aspects.
First, \textsc{3dm-merge} iterates over all elements of the change set and
intermingles the activities of processing content and PCS triples.
\textsc{spork-3dm} on the other hand iterates only over the PCS triples and
separates the processing of content tuples and PCS triples, which makes it
possible to reason about the merging of structure and content separately. Due
to how far removed the merging of content is from the merging of PCS triples,
it is perfectly viable for an implementation of \textsc{spork-3dm} to defer the
merging of content until building an AST from the merged change set. Second,
there is a key functional difference in how the algorithms discover
inconsistencies. \textsc{3dm-merge} finds at most one inconsistent element per
iteration of the primary loop, while \textsc{spork-3dm} finds all of them. With
the original algorithm, hard inconsistency detection sometimes becomes
non-deterministic when the same PCS triple is involved in inconsistencies with
many other triples.

\begin{table*}[t]
    \caption{The effect of consecutive invocations of the
        \lstinline{removeSoftPcsInconsistencies} function in the core loop of the
        \lstinline{merge} function starting from the raw merge from
        \autoref{tab:3dm:pcs_sets}.  For brevity, the table illustrates the
        invocations until all hard inconsistencies have been discovered and the
        final merge is achieved, as subsequent invocations have no effect.}
    \label{tab:spork:dry_run}
    \centering
    \begin{tabular}{l|lll}
        pcs (input)      & inconsistencies                                & removals         & hard inconsistencies \\
        \toprule
        $(01,\dashv,02)$ & $(01,\dashv,12),(12,\dashv,02),(12,02,\vdash)$ & $(01,\dashv,02)$ & - \\
        $(12,\dashv,02)$ & $(01,02,03)$                                   & $(01,02,03)$     & - \\
        $(01,03,15)$     & $(01,03,\vdash),(01,03,25)$                    & $(01,03,\vdash)$ & $(01,03,25)$ \\
        $(01,15,\vdash)$ & $(01,25,\vdash)$                               & -                & $(01,25,\vdash)$ \\
    \end{tabular}
\end{table*}

\subsection{Building the AST and Handling Conflicts}\label{sec:spork:conflicts}

When the change set has been merged, it must be converted back into an
AST. This is achieved by traversing the PCS set and inserting visited
nodes and their contents into an AST. Note that excess structural information
is discarded when building the AST, such as the virtual root ($\bot$), start
($\dashv$) and end ($\vdash$), as well as the intermediate virtual nodes
discussed in \autoref{sec:spoon_to_pcs}.

Given a PCS, traversal left and right within the child list amounts to
finding another PCS with the same parent and one matching child node, but where
said child node's position is different. The child list of any node $x$ starts
with a PCS where $x$ is the parent, and $\dashv$ is the predecessor. These
traversal rules are summarized in \autoref{tab:spork:pcs_traversal}. Also
recall that the content of an arbitrary node $v$ is represented by all content
tuples $c(v, m)$.

In a child list without conflicts, there is always precisely one PCS
matching each traversal pattern from some starting point, and each node of
that PCS has precisely one content tuple. Consider again the merged PCS set
in \autoref{tab:3dm:pcs_sets}. A traversal always begins from the start of the
virtual root's child list, which according to the traversal rules is
$(\bot,\dashv,01)$. It is a simple matter to traverse this child list to find
that $01$ is the concrete root of the tree, and similarly that the first two
children of $01$ are $12$ and $03$. Upon encountering each of these nodes in
the traversal, their contents can be extracted from the merged content set of
\autoref{tab:spork:content_merge}. However, the PCS set is clearly not
consistent, as there are two $PCS$ triples matching the right traversal pattern
from $(01,12,03)$. This hard inconsistency indicates a conflict in the merge,
the handling of which is described in
\autoref{sec:spork:insert_insert_conflicts}.

\begin{table}[t]
    \caption{Traversal rules for the PCS structure, starting from an
        arbitrary initial PCS $(x, y, z)$. Traversing in a given direction amounts
        to finding a PCS matching a specific pattern, where $?$ is an unknown
        node.}
    \label{tab:spork:pcs_traversal}
    \centering
    \begin{tabular}{l|l}
        Direction & PCS pattern \\
        \toprule
        Left & $(x,?,y)$ \\
        Right & $(x,z,?)$ \\
        Into $y$'s child list & $(y,\dashv,?)$ \\
        Into $z$'s child list & $(z,\dashv,?)$ \\
    \end{tabular}
\end{table}

\begin{table}[t]
    \centering
    \caption{Content merge of the merge scenario of \autoref{fig:three_way_tree_matching}}
    \label{tab:spork:content_merge}
    \begin{tabular}{c}
        $c(01,"sum"), c(02,"a"), c(03,"b"), c(12, -), c(15, 1), c(25, "c")$
    \end{tabular}
\end{table}

\subsubsection{Insert/insert conflicts}\label{sec:spork:insert_insert_conflicts}

An insert/insert conflict occurs when both revisions insert one or more nodes in
the same position in the AST. The hard inconsistencies in
\autoref{tab:3dm:pcs_sets} are caused by such a conflict, namely the
insertion of $int_{15}(1)$ in the left revision and $ref_{25}(c)$ in the right
revision (see \autoref{fig:three_way_tree_matching}). The inconsistent elements
show a typical pattern for an insert/insert conflict, namely that the conflict
starts with the successor inconsistency between $(01,03,15)$ and $(01,03,25)$,
and ends with the predecessor inconsistency between $(01,15,\vdash)$ and
$(01,25,\vdash)$. This yields two possible paths from node $03$ to the virtual
end $\vdash$, either through node $15$ or through node $25$.

\spork handles insert/insert conflicts by traversing both paths through the
child list and collecting the nodes of both sides of the conflict. These are
inserted into a conflict node, which in this case contains $15$ from the
left revision and $25$ from the right revision. The full AST represented by the
merge in \autoref{tab:3dm:pcs_sets} can then be built, resulting in the tree
shown in \autoref{fig:spork:merged_tree}.

\begin{figure}[t]
    \begin{center}
        \begin{tikzpicture}[sibling distance=7em, every node/.style={rectangle, draw=black, align=center, thick}, every label/.style={rectangle,draw=none}]
            \small

            \begin{scope}
                \tikzstyle{edge from parent}=[draw,black,thick,->]
                \node (call) {call$_{01}$(sum)}
                child {node (uop) {uop$_{12}$(-)}
                    child {node (lhs) {ref$_{02}$(a)}}
                }
                child {node (mid) {ref$_{03}$(b)}}
                child {node (rhs) {conflict \\left: [$int_{15}(1)$]\\right: [$ref_{25}(c)$]}}
                ;
            \end{scope}
        \end{tikzpicture}
    \end{center}
    \caption{AST built from the PCS merge in \autoref{tab:3dm:pcs_sets}
    and content set in \autoref{tab:spork:content_merge}}
    \label{fig:spork:merged_tree}
\end{figure}

\subsubsection{Delete/delete conflicts}\label{sec:spork:delete_delete}

A delete/delete conflict occurs when the left and right revisions delete
adjacent nodes. For example, consider that the base revision has a node $p$
with a child list $ab$, and that the left revision deletes $b$ while the right
revision deletes $a$. Omitting the parent node, the base revision's PCS child
list is then $(\dashv,a),(a,b),(b,\vdash)$, the left revision's is
$(\dashv,a),(a,\vdash)$, and the right revision's is $(\dashv,b),(b,\vdash)$.
When merging these child lists, the first and last elements are inconsistent
across the left and right revisions. However, as $(\dashv,a)$ and $(b,\vdash)$
are both in the base revision, both inconsistencies are soft and can be
eliminated, resulting in the consistent but clearly disjoint child list
$(\dashv,b),(a,\vdash)$. This shows that the consistency criteria are not
strong enough to guarantee that a consistent change set encodes a well-formed
tree.

Disjoint child lists are trivial to detect, as they always result in one PCS
with a successor (e.g. $b$ in $pcs(p, \dashv, b)$) that never appears as a
predecessor in the same child list. However, \spork currently cannot recover all
conflicting AST nodes due to the fact that parts of the conflict have already been
removed from the change set. Instead, it falls back on a line-based merge of
the textual representations of the subtrees rooted in parent node $p$. We refer
to this as the \emph{local fallback}.
\autoref{tab:spork:delete_delete_conflict} shows an example merge scenario of a
code snippet where the delete/delete conflict already discussed occurs in the
argument list of a method call. The resulting merge conflict can be seen in
\autoref{fig:spork:local_fallback_conflict}, along with the fully line-based
merge for comparison. \spork expresses conflicts in the same way that
\textsc{git}'s default merge tool does, with so-called \emph{conflict
hunks}. Each hunk starts with left-facing arrows ($<$) followed by the left
revision's part and ends with right-facing arrows ($>$) preceded by the right
revision's part, the two parts being demarcated by a line of equals signs
($=$). We use the terms conflict and conflict hunk interchangeably.

Note that the lines merged by the local fallback are not necessarily the exact
lines of the original source files, but the textual representation of the
conflicting subtrees. Thus, although less granular than structured conflict
handling, the local fallback is still significantly more granular than a
line-based merge of the entire file. A conflict node is then inserted into the
AST containing the line-based merge, which is printed as-is during
pretty-printing.

\begin{table}[t]
    \caption{The left, base and right revisions of a line with a delete/delete conflict, and the textual representations of the subtrees with conflicting child lists}
    \label{tab:spork:delete_delete_conflict}
    \centering
    \begin{tabular}{l|ccc}
                         & Left          & Base            & Right \\
        \toprule \\
        Line             & abs(sum(a)) & abs(sum(a,b)) & abs(sum(b)) \\
        Conflict subtree & sum(a)       & sum(a,b)       & sum(b) \\
    \end{tabular}
\end{table}

\begin{figure}[t]
    \centering

    \begin{subfigure}[t]{0.45\linewidth}
        \centering
        \begin{tabular}{c}
            \begin{lstlisting}
abs(
<<<<<<< left
sum(a)
=======
sum(b)
>>>>>>> right
)
            \end{lstlisting}
        \end{tabular}
        \caption{Structured merge with local fallback}
    \end{subfigure}
    ~
    \begin{subfigure}[t]{0.45\linewidth}
        \centering
        \begin{tabular}{c}
            \begin{lstlisting}
<<<<<<< left
abs(sum(a))
=======
abs(sum(b))
>>>>>>> right
            \end{lstlisting}
        \end{tabular}
        \caption{Fully line-based merge}
    \end{subfigure}
    \caption{Merge produced by \spork's local fallback activating on the merge in \autoref{tab:spork:delete_delete_conflict} and the merge of a traditional line-based merge tool}
    \label{fig:spork:local_fallback_conflict}
\end{figure}

\subsubsection{Insert/delete conflicts}\label{sec:spork:insert_delete}

An insert/delete conflict occurs when one revision inserts a node where
another revision deletes a node. For example, assume that a node $p$ has a child list
$a$ in the base revision, that the left revision deletes $a$, and the right
revision inserts $b$ after $a$. Omitting $p$, the base child list is then
$(\dashv,a),(a,\vdash)$, the left is $(\dashv,\vdash)$ and the right is
$(\dashv,a),(a,b),(b,\vdash)$. Removing all soft inconsistencies from the raw
merge results in the child list $(\dashv,\vdash),(a,b),(b,\vdash)$. The first
and last elements are in a hard predecessor inconsistency, and the middle
element is disjoint from the start of the child list.

As with the delete/delete conflict discussed in
\autoref{sec:spork:delete_delete}, it is difficult to retrieve the left and
right sides of the conflict due to the fact that part of the conflict is not
present in the final change set. \spork therefore resorts to the local fallback
described in \autoref{sec:spork:delete_delete} when discovering an
insert/delete conflict.

\subsubsection{Move conflicts}\label{sec:spork:move_conflicts}

Move conflicts are theoretically and practically troublesome. Intuitively, a
move conflict occurs when the left and right revisions both manipulate the same
node or the context around it such that the node ends up in two different
positions in the computed merge. A pure move conflict involves the left and
right revisions moving the same node to two different locations. Moves may also
conflict with insertions (move/insert) or deletions (move/delete) at the source
or destination sites. For example, assume that a node $p$ has a child list $a$
in the base revision, that the left revision moves $a$ to another child list,
and the right revision inserts $b$ after $a$. This is a move/insert conflict,
and in fact causes the child list of $p$ to take exactly the same form as with
the insert/delete conflict discussed in \autoref{sec:spork:insert_delete}. The
difference in this case is that $a$ also exists in some other child list.

There are two high-level variations of move conflicts that present differing
difficulties to handle. An \emph{inter-parent} move conflict occurs when a node
$v$ is involved in a hard parent inconsistency, that is to say, has two
different parents. Parent inconsistencies are easy to detect, and can only be
caused by move conflicts, such as the one described above where $a$ ends up in
two child lists. \spork handles such conflicts by recursively classmapping all
nodes in the subtrees rooted in $v$ to themselves, and then restarting the
merge. As the nodes in the revisions of $v$ are no longer classmapped to each
other, they are considered different nodes by \textsc{spork-3dm}, which
effectively turns moves into insertions and deletions. This may resolve the
conflict entirely, or result in non-move conflicts that are easier to deal
with.

An \emph{intra-parent} move conflict occurs when a node $v$ appears in two
places in the same child list. These conflicts are hard to identify as move
conflicts as there is no single feature that distinguishes the resulting
inconsistencies from those caused by other conflicts, delete conflicts in
particular. Therefore, determining which nodes to classmap to themselves is
difficult, and so \spork resorts to the local fallback instead.

\subsubsection{Conflict handlers}\label{sec:spork:structural_conflict_handlers}

In some cases, conflicts can be automatically resolved. In the case of
structural conflicts where \spork successfully extracts all AST nodes that
take part in a conflict, \spork invokes two \emph{structural conflict handlers}. The first of these resolves conflicts
involving ambiguous ordering of method declarations\footnote{So-called
\emph{ordering conflicts}}, by inserting them in sorted order. The second
handler resolves conflicts where one of the conflicting sides
is empty by picking the non-empty side, optimistically: if a handler can resolve the
conflict, the conflict node is replaced by a concrete node.

\spork also defines \emph{content conflict handlers}. When
building the AST and attempting to set the content of a node for which multiple
content tuples are found, \spork invokes the content conflict handlers one by
one until the conflict is resolved, or there are no more conflict handlers.
Most content conflict handlers are highly dependent on \textsc{spoon}
implementation details, and therefore fall outside of the scope of this
paper, and we refer the reader to the implementation for
details\footnote{\sporkrepo}.

\subsubsection{Duplicated Type Member Elimination}

The built \textsc{spoon} AST is then subjected to \emph{duplicated type member
elimination}. In cases where both the left and right revisions add
non-identical versions of the same type member, a failure to match these
results in type member duplication in the merge process. For example, a class
may end up with two methods with the same signature, or two fields with the
same name, making for semantic conflicts. As \textsc{3dm-merge} knows nothing
of the semantics of \textsc{java}, a duplicated type member will not seem
problematic to it. To address such issues, \spork searches the merged
\textsc{spoon} AST for duplicated type members, and re-executes the entire
merge process for any pair it can find, using an empty node as the base
revision\footnote{This effectively makes it a two-way merge of the type
members}. The reason why duplicated type member elimination is performed at such a
late stage is a matter of implementation convenience; it is trivial to find
duplicated type members in the merged \textsc{spoon} tree, while doing so in
any of the earlier stages is much harder.

\subsection{Pretty-printing}\label{sec:spork:pretty_printing}

\begin{figure}[t]
    \centering
    \begin{tabular}{c}
        \begin{lstlisting}
sum(-a, b,
<<<<<<< left
1
=======
b
<<<<<<< right
)
        \end{lstlisting}
    \end{tabular}
    \caption{Pretty-printed output of the AST in \autoref{fig:spork:merged_tree}}
    \label{fig:spork:pretty_print}
\end{figure}

\spork uses \textsc{spoon}'s default pretty-printer to produce the
final result of the merge, namely a \textsc{java} source file. There are
however two significant extensions to the printer in \spork.

The first extension is reusing the original source code for subtrees that
originate from a single revision, in effect performing a copy-paste of a
subtree's original source code. We refer to this as \emph{high-fidelity
pretty-printing}, and it allows \spork to retain most of the formatting from
the file revisions. High-fidelity pretty-printing of a merged AST is currently
limited to type members and comments due to complications at more granular
levels when adjacent elements stem from different revisions. For
example, \spork can directly reuse the source code of a method declaration that
it determines stems from a single revision. However, if \spork finds that a method
declaration is composed of elements from multiple revisions, it currently cannot
perform high-fidelity pretty-printing of individual child elements of that
method, such as method parameters and statements. Either the entire method is
high-fidelity pretty-printed, or none of it is. For printing of more granular
elements whose parent elements cannot be printed with high-fidelity
pretty-printing, \spork uses \spoon's default pretty-printer, only taking the
original indentation into account. All other formatting is fixed for
each type of AST node. For example, the last method parameter in a parameter
list is always immediately followed by a closing parenthesis, while all
non-last parameters are followed by a comma and a space. We refer to this as
\emph{low-fidelity pretty-printing}.

The second extension is printing of conflicts. \spork uses high-fidelity
pretty-printing to print both sides of a structural conflict, or directly prints
the contents of a conflict node produced by the local fallback (see
\autoref{sec:spork:delete_delete}). Note that the limitation of what elements
can be printed with high-fidelity pretty-printing do not apply here, as the
nodes of any given side of a conflict by definition stem from the same
revision. The pretty-printed output of the running example is shown in
\autoref{fig:spork:pretty_print}, containing such a conflict.

\section{Experiment Methodology}\label{sec:meth}

This section presents the methodology we use for the evaluation of \spork. We
compare \spork against \jdime~\cite{apel2012structured}, a state-of-the-art
structured merge tool for \textsc{java}, and
\automergeptm~\cite{Zhu2019}, a merge tool that builds upon \jdime with an
enhanced tree matching algorithm. We assess \spork in regards to conflicts,
running time and formatting preservation.

\subsection{State-of-the-art of Structured Merge for Java}

We select the state-of-the-art tools for structured merge in Java as follows.
First, the state-of-the-art tool does fully-structured merge on ASTs,
which allows for an apples-to-apples comparison with \spork. Second, there is
a publicly available code base, which is crucial for reproducibility. Third,
it works on real-world and non-trivial merge scenarios.

We have evaluated the tools in \autoref{tab:meth:considered_merge_tools}
according to these criteria. Only two tools fulfill all criteria:
\jdime~\cite{apel2012structured} and \automergeptm~\cite{Zhu2019}. \jdime is
fully structured and built on top of the \textsc{extendj} compiler framework
and is able to merge real-world merge scenarios. \jdime has been extensively
used in subsequent research, incl.
\cite{Le_enich_2014,Lesenich2017,cavalcanti2019impact,Zhu2018,Zhu2019}.
\automergeptm is mostly the same tool as \jdime, but with an enhanced
tree matching algorithm.

\begin{table}[t]
    \setlength{\tabcolsep}{4pt}
    \centering

    \begin{threeparttable}
        \caption{Potential Merge Tools for Evaluation}
        \label{tab:meth:considered_merge_tools}
        \begin{tabular}{lp{3cm}ll}
            Tool                                              & Key features                                            & Comments \\
            \midrule
            \jdime~\cite{apel2012structured}\tnote{1}         & Structured, AST-based                                   & Included \\
            \hline
            \automergeptm~\cite{Zhu2019}\tnote{2}             & Structured, built on \jdime with improved tree matching & Included \\
            \hline
            \textsc{jfstmerge}~\cite{Cavalcanti2017}\tnote{3} & Semistructured                                          & Excluded \\
            \hline
            \textsc{intellimerge}~\cite{Shen2019}\tnote{4}    & Semistructured, graph-based                             & Excluded \\
            \hline
            Envision IDE~\cite{Asenov2017}\tnote{5}           & Line-based merge on textual AST                         & Excluded \\
            \bottomrule
        \end{tabular}

        \scriptsize
        \begin{tablenotes}
            \item[1]\url{https://github.com/se-sic/jdime} \\
            \item[2]\url{https://github.com/thufv/automerge-ptm} \\
            \item[3]\url{https://github.com/guilhermejccavalcanti/jFSTMerge} \\
            \item[4]\url{https://github.com/symbolk/intellimerge} \\
            \item[5]\url{https://github.com/dimitar-asenov/Envision}
        \end{tablenotes}
    \end{threeparttable}
\end{table}

\subsection{Research Questions}\label{sec:meth:rqs}

The evaluation is structured around the following research questions.

\begin{itemize}
\item    \textbf{RQ1: How does \spork compare to \jdime and \automergeptm in terms of amounts of conflicts and amounts of conflicting lines?} Merge conflict prevalence is a key aspect of a merge tool. The reduction of the number of merge conflicts is a primary advantage of structured
merge. Furthermore, conflict size is an indicator that developers use to estimate the conflict difficulty, increasing size being associated with increasing difficulty~\cite{McKee2017}.

\item    \textbf{RQ2: How does \spork compare to \jdime and \automergeptm in terms of running time?} Running time is an important aspect for any software engineering tool that is used in an interactive
computing environment. The user has to wait for the merge tool to run to completion before being able to proceed, meaning that all else being equal, a faster tool is preferable to a slower one. 

\item    \textbf{RQ3: How does \spork compare to \jdime and \automergeptm in terms of preserving source code formatting?} 
Due to operating on an abstract representation of source code, a structured
merge always concludes in a pretty-printing step. If the printer does not
attempt to recreate the formatting of the input, it may fundamentally alter
it~\cite{Waddington2007,Savchenko2019}. We argue that a merge tool should not
alter formatting at all.

\end{itemize}

\subsection{Dataset}\label{sec:meth:dataset}
We select projects for the experiments from the \textsc{reaper} dataset of
\textsc{github} repositories~\cite{Munaiah2017}. This dataset consists of 1.8
million \textsc{github} projects that are scored with respect to a variety of
indicators of a well-engineered project. These indicators include the use of CI
and unit tests, the amount of documentation and amount of core contributors.
The \textsc{reaper} dataset has  been used in other merge
studies~\cite{cavalcanti2019impact,Owhadi-Kareshk2019a}.

\subsubsection{Filtering projects}\label{sec:meth:filtering_projects}
We select projects in \textsc{reaper} that use \textsc{java}, are classified as
well-engineered, have more than 50
stars and a minimum of 2 core
contributors. A total of 1174 projects fulfill all of the
criteria. We further filter out projects that are forks.

We perform a last filtering to find projects that build using \textsc{maven} in
our test environment. We look for a \texttt{pom.xml} file in the latest
commit\footnote{As of the 10th of May 2020} of the default branch, indicating
the use of \textsc{maven}. If there is such a file, the project is cloned and
built with \textsc{maven}. If the build succeeds within at most 5 minutes, the
project is added to a list of candidate projects. We use two of these projects
for testing purposes during the development of \spork, and we exclude them from
the evaluation. 

This selection process leads to a list of 359 candidate projects that fulfill
all criteria. This list is available in the online
appendix\footnote{\sporkappendix}.

\subsubsection{Filtering merge commits}\label{sec:meth:filtering_merge_commits}
Our experiments require the base commit of each merge scenario to be located.
We use \textsc{git-log} to find merge commits, and \textsc{git-merge-base} to
find the merge base. In cases where a commit history has a criss-cross
pattern\footnote{\url{https://git-scm.com/docs/git-merge-base}}, there are
multiple possible merge bases. The merge base is then said to be
\emph{ambiguous}. Merge commits with ambiguous merge bases are excluded from the
dataset as they complicate merge replay.

As noted by Cavalcanti et al.~\cite{cavalcanti2019impact}, most merges have no overlap between
the files edited by the left and right revisions, making the merge resolution
trivial: simply pick the edited file. \textsc{git} only invokes a merge tool
when the same file has been edited in both revisions. The merge commits are
therefore filtered to include only commits where at least one \textsc{java}
source file is edited in both the left and right revisions.

We also filter out merge commits for which at least one of the revisions do not
build using \textsc{maven}. If the project builds, we can be certain that it is
syntactically valid, which is important as syntactically invalid files can cause
unexpected behavior in structured merge tools. Finally, as some projects have
thousands of merge commits, while others have as little as 1, we limit the
amount of merge commits per project to 15 to avoid the larger projects being
overrepresented.

We extract a total of 890 real-world merge scenarios from 119 different
projects, consisting of a total of 1740 file merges. We observe a great deal of
variety in project domains, such as the \textsc{mage} game engine, the
\textsc{corenlp} natural language processing library, the \textsc{assertj}
assertions library, the \textsc{singularity} platform-as-a-service and the
\textsc{chronicle-map} in-memory database. The popularity and sizes of projects
also vary greatly.  Project sizes range from 1106 to 1782052 lines of code, with
a median of 24306.  The amount of GitHub stars ranges from 58 to 21720, with a
median of 341. The amount of core contributors ranges from 1 to 40, with a
median of 6. Finally, the amount of merge scenarios extracted from each project
ranges from 1 to 15, with a median of 7, and the amount of file merges ranges
from 1 to 122, with a median of 12. Note that the project metadata was collected
on August 12th 2020 while the \textsc{reaper} dataset is from 2017, meaning that
there are some discrepancies between the two. The full list of project
statistics is available in the online appendix\footnote{\sporkappendix}.

\subsection{Experiment Protocol}\label{sec:meth:file_merge}
We design an experiment protocol similar to that of Shen et
al.~\cite{Shen2019}. For each merge scenario, the individual file merges are
extracted. This involves finding all revisions of a merged file, including the
one actually committed by the developer, which we refer to as the
\emph{expected} revision of the file merge. We use \textsc{git}'s merge
functionality to locate the revisions.

\spork and \jdime are then applied in turn to the base, left and right
revisions of each file merge. We refer to the merged file produced by a merge
tool as the \emph{replayed} revision for that file merge and merge tool.

In order to assess RQ1, we scan each replayed revision for conflict
hunks\footnote{Recall the definition of conflict hunks from
\autoref{sec:spork:delete_delete}}, and record the amount of hunks as well as
the total amount of lines contained in them. 
In general, it is easier for developers to deal with conflicts if they are
as few and small as possible~\cite{Menezes2018,Nelson2019,vale2021challenges}, meaning that
minimizing conflict quantities and sizes is desirable. To make sure that we
only analyze merge conflicts produced by the tools under test, any file merges
in which the base, left or right revisions contain conflict markers are
excluded from the experiment.  For all comparisons, file merges where at least
one tool fails to produce a non-empty merged file are excluded. This also
applies to subsequent research questions.

To address RQ2, the execution of each file merge is timed 10x per file
file, with the running time measured as the wall time from the moment of
invoking the merge tool to the moment it exits. Each execution is a cold start,
meaning that the JVM is not allowed to warm up. A timeout is set to 300 seconds
per file merge, after which the merge is forcibly aborted.

To address RQ3, we measure formatting preservation with the expected revision
as the ground truth for correctly formatted output. In order to determine how
closely the replayed revision resembles the expected revision, we compare them
with two metrics: a \emph{\linediff} computed with
\textsc{git}\footnote{\url{https://git-scm.com/docs/git-diff}}, and a
\emph{\chardiff} computed with the \textsc{python} standard library module
\textsc{difflib}~\footnote{\url{https://docs.python.org/3.8/library/difflib.html}}.
We refer to the sum of insertions and deletions of lines and characters as the
\emph{\linediff size} and \emph{\chardiff size}, respectively. Poor formatting
preservation increases the diff size.

For all RQs, we illustrate the behavior of \spork with case studies. Those case
studies are real merge scenarios taken from our dataset. They are selected
manually, with the goal of highlighting advantages and drawbacks of \spork.

\subsection{Statistical Tests}
\label{sec:meth:stats}

All measurements provide us with paired ordinal data (conflict sizes, conflict
quantities, diff sizes and running times) for each merge scenario (one measure
for \jdime, one for \spork, and one for \automergeptm). As the tools might
fail, we exclude all measures from merge scenarios where at least one tool
fails to produce a merge.

To statistically assess the differences between the tools, we first start by a
Friedman test with the null hypothesis that the measures for all tools are the
same. In case of a significant p-value on the significance level of $\alpha =
0.05$, the null hypothesis is rejected and we then proceed to two post-hoc
tests to compare the groups \spork vs \jdime and \spork vs \automergeptm. We
use a two-sided Wilcoxon signed-rank test, along with the matched-pairs
rank-biserial correlation (RBC) as effect size~\cite{Kerby2014,tomczak2014need}
for each post-hoc test. The resulting p-values are then corrected using a
Holm-Bonferroni correction. Finally, we assess their  significance using
$\alpha = 0.05$. We use the implementation provided by the \textsc{python}
package \textsc{pingouin}\footnote{\url{https://pingouin-stats.org/}} version
0.4.0 to perform the tests and calculate effect sizes. In our results, a
negative RBC indicates that \spork's values in the given test are smaller than
\jdime's or \automergeptm, while a positive RBC indicates the opposite.

\subsection{Experiment Environment}\label{sec:meth:system}
The test environment hardware consists of a Ryzen 5900X, 32 GiB of RAM
@3600 MHz and a SATA SSD with read and write speeds of 500 MB/s. The test
environment runs \textsc{archlinux} with kernel 5.13.9, \textsc{openjdk}
1.8.0u292 and \textsc{cpython} 3.8.2. We build \jdime from source using commit
100aeece. We build \automergeptm from source using commit e73038b5. We use
\spork release v0.5.1. For further information, we refer the reader to the
online appendix\footnote{\sporkappendix}.

\section{Experiment Results}\label{sec:res}

This section presents the results from the experiments.
\autoref{sec:res:conflicts} presents results on sizes and quantities of
conflicts, \autoref{sec:res:running_times} presents results on running times
and \autoref{sec:res:formatting} presents results on formatting preservation.

\subsection{RQ1: Quantity and Size of Conflicts}
\label{sec:res:conflicts}
\subsubsection{Amount of conflict hunks}\label{sec:res:conflict_hunk_quantity}
\begin{figure}
    \centering
    \resizebox{\linewidth}{!}{\begingroup \makeatletter 
\makeatother \endgroup  }
    \caption{RQ1: Histogram of conflict hunk quantities per file for \spork,
        \jdime and \automergeptm. Lower is better. Each histogram bin contains the
        frequency of values in the range $[L, R)$, where $L$ and $R$ are the values to
        the left and right of the bin, respectively.}
    \label{fig:res:conflict_quantity_histogram}
\end{figure}

We first measure and  compare the amount of conflict hunks per file. As
noted in \autoref{sec:meth:file_merge}, fewer and smaller conflicts is
generally better. However, there are cases where fewer or smaller conflicts are due to a poor
merge, such as when truly conflicting edits are not detected as such, or when
a conflict is not intuitively represented. This is
discussed in the illustrative case studies below.

As noted in \autoref{sec:meth:file_merge}, we filter out file merges
where at least one merge tool fails to produce a non-empty file merge, as
it is then not possible to fairly compare the results. There are three
separate cases that can occur: the merge tool can crash, exceed the time
limit of 300 seconds or produce an empty merged file. An empty or
non-existing file cannot be used for making comparisons of our chosen
metrics, and must therefore be excluded. \autoref{tab:res:merge_failures}
shows a breakdown of merge failures across the tools. \spork has the most
amount of crashes, but exhibits none of the other kinds of failures. \jdime
has the smallest amount of failures in total, but it also exhibits the largest
amount of timeouts. \automergeptm has the largest amount of failures in total.
It is the only tool to occasionally produce an empty merged file,
accounting for most of its failures, but it also exhibits both the other
kinds of failures. There is little overlap between the file merges where the
tools fail, with the 94 merge failures occurring across 83 unique file merges,
constituting 4.77\% of the total of 1740 file merges.

\begin{table}[t]
  \renewcommand{\arraystretch}{1.7}
  \centering
  \begin{tabular}{lrrrr}
      \textbf{Tool} & \textbf{Timeouts} & \textbf{Crashes} & \textbf{Empty file} & \textbf{Total} \\
      \hline
      \spork        & 0                 & 34               & 0                   & 34 \\
      \jdime        & 16                & 7                & 0                   & 23 \\
      \automergeptm & 7                 & 11               & 19                  & 37 \\
  \end{tabular}
  \caption{Summary of merge failures.}
  \label{tab:res:merge_failures}
\end{table}

Out of the 1740 file merges in the benchmark, there are 255 file merges
for which all of \spork, \jdime and \automergeptm produce a non-empty merge
file, and at least one tool encounters a conflict. \spork signals conflicts
in 125 of the merges and produces a total of 227 conflict hunks. \jdime
signals conflicts in 191 of the merges and produces a total of 376 conflict
hunks. \automergeptm signals conflicts in 145 of the merges and produces a
total of 245 conflict hunks.  Overall, \spork produces 151 conflict hunks
fewer than \jdime (40\% reduction), and 18 fewer than \automergeptm (7\%
reduction).

\autoref{fig:res:conflict_quantity_histogram} shows the histogram of the
distribution of conflict hunk quantities for \spork, \jdime and \automergeptm.
While \jdime is clearly at a disadvantage (more files with 2 or more conflict
hunks), the distributions for \spork and \automergeptm look largely similar.

We use a Friedman test to determine if further analysis of the results is
relevant, with the null hypothesis that the results from the different tools are
the same. The test yields a p-value of 2.40e-13, so we reject the null hypothesis
and proceed with further analyses.

We use a two-sided Wilcoxon signed-rank test to test the following hypothesis:

\begin{quote}
    $H_0^1$: \spork and \jdime produce the same amounts of conflict hunks

    $H_a^1$: \spork and \jdime do not produce the same amounts of conflicts
\end{quote}

\noindent
The test yields a p-value of 7.98e-8, and we therefore accept the alternative
hypothesis that the tools produce differing amounts of conflict hunks. The RBC
is -0.423, indicating that \spork produces fewer hunks than \jdime.

We use a two-sided Wilcoxon signed-rank test to test the following hypothesis:

\begin{quote}
    $H_0^2$: \spork and \automergeptm produce the same amounts of conflict hunks

    $H_a^2$: \spork and \automergeptm do not produce the same amounts of conflicts
\end{quote}

\noindent
The test yields a p-value of 0.269, so we cannot reject the null hypothesis that
the tools produce the same amounts of conflicts.

\subsubsection{Amount of conflicting lines}\label{sec:res:conflict_hunk_size}

\begin{figure}
    \centering
    \resizebox{\linewidth}{!}{\begingroup \makeatletter 
\makeatother \endgroup  }
    \caption{Histogram of conflict sizes for \spork,
        \jdime and \automergeptm per file merge. Lower is better. Each histogram bin contains the
        frequency of values in the range $[L, R)$, where $L$ and $R$ are the values to
        the left and right of the bin, respectively.}
    \label{fig:res:conflict_size_histogram}
\end{figure}

We now consider the amount of conflicting lines per file merge, which we refer
to as the \emph{conflict size}. We consider here the same 255 file merges as in
\autoref{sec:res:conflict_hunk_quantity}; file merges where all tools produce a
non-empty merged file and at least one tool produces a conflict hunk. We
measure the conflict size of a file as the sum of all lines in all conflict
hunks (see \autoref{sec:spork:delete_delete}), which is a proxy to the effort
spent by developers to resolve conflicts~\cite{Nelson2019,Menezes2018,vale2021challenges}.

\autoref{fig:res:conflict_size_histogram} shows a histogram of conflict sizes
for \spork, \jdime and \automergeptm. The first bin refers to cases where the
merge is fully successful, containing no conflict\footnote{Note that it is
    identical to that of the conflict quantity histogram in
\autoref{fig:res:conflict_quantity_histogram}.} In this bin, \spork outperforms
both \jdime and \automergeptm. Looking at the rightmost bin of the figure,
\jdime and \automergeptm produce more files with large conflict sizes.  For all
tools, there are outliers, meaning that a small amount of conflicts account for
the majority of conflicting lines.  This is the primary explanation of \spork's
improvement: \automergeptm and \jdime produce more abnormally large conflicts
than \spork. In particular, \spork produces files with conflict sizes at or
above 20 lines of code in 24 merges, making for a reduction by 54\% compared to
\jdime's 52 cases, and by 47\% compared to \automergeptm's 45 cases. In the
middle of the distribution, the interpretation is not clear-cut, but accounts
for significantly fewer data points than the extrema.

In the case of \jdime and \automergeptm, abnormally large conflicts are often
caused by failures to match renamed elements to each other, which is
exemplified with a method rename in case study \caserename discussed below.
Regarding \spork's large conflicts, they are often caused
by the local-fallback activating on the body of a class, causing most of the
file to be merged with a line-based merge.

Let us now aggregate these results.
Over the 255 file merges,
\spork produces a total of 2446 conflicting lines, \jdime produces a total of
13975 conflicting lines, and \automergeptm produces a total of 6635 conflicting
lines. 
\spork improves upon \automergeptm, second best by this metric, by 63\%.

We use a Friedman test to determine if further analysis is necessary,
with the null hypothesis that the results from the different tools are the
same. The test yields a p-value of 2.23e-13, so we reject the null hypothesis and
proceed with further analyses.

We use a two-sided Wilcoxon signed-rank test to test the following hypothesis:

\begin{quote}
    $H_0^3$: \spork and \jdime produce equal amounts of conflicting lines
    
    $H_a^3$: \spork and \jdime do not produce equal amounts of conflicting lines
\end{quote}

\noindent
The test yields a p-value of 4.47e-4, and we therefore accept the
alternative hypothesis that the tools do not produce equally large conflicts.
The effect size RBC is -0.280, indicating that \spork produces fewer conflicting
lines than \jdime.

We also use a two-sided Wilcoxon signed-rank test to test the following
hypothesis:

\begin{quote}
    $H_0^4$: \spork and \automergeptm produce equal amounts of conflicting lines
    
    $H_a^4$: \spork and \automergeptm do not produce equal amounts of conflicting lines
\end{quote}

\noindent
The test yields a p-value of 0.441, so we cannot reject the null
hypothesis that the tools produce equal amounts of conflicting lines.

\emph{Conflict case studies}\label{sec:res:conflict_behavior}

\begin{figure}[t]
\scriptsize
\centering
    \begin{subfigure}[t]{0.99\linewidth}
        \centering
        \begin{tabular}{c}
            \begin{lstlisting}[linewidth=0.95\textwidth,basicstyle=\scriptsize,language=java,numbers=none]
/**
 * Copyright 2009-2019 ...
 *\end{lstlisting}
        \end{tabular}
        \caption{Left revision}
    \end{subfigure}
    \par\bigskip

    \begin{subfigure}[t]{0.99\linewidth}
        \centering
            \begin{tabular}{c}
                \begin{lstlisting}[linewidth=0.95\textwidth,basicstyle=\scriptsize,language=java,numbers=none]
/**
 * Copyright 2009-2016 ...
 *\end{lstlisting}
        \end{tabular}
        \caption{Base revision}
    \end{subfigure}
    \par\bigskip

    \begin{subfigure}[t]{0.99\linewidth}
        \centering
        \begin{tabular}{c}
            \begin{lstlisting}[linewidth=0.95\textwidth,basicstyle=\scriptsize,language=java,numbers=none]
/**
 * Copyright 2009-2020 ...
 *\end{lstlisting}
        \end{tabular}
        \caption{Right revision}
    \end{subfigure}
    \par\bigskip

    \begin{subfigure}[t]{0.99\linewidth}
        \centering
        \begin{tabular}{c}
            \begin{lstlisting}[linewidth=0.95\textwidth,basicstyle=\scriptsize,language=java,numbers=none]
/**
<<<<<<< LEFT
 *    Copyright 2009-2019 ...
=======
 *    Copyright 2009-2020 ...
>>>>>>> RIGHT
 *\end{lstlisting}
        \end{tabular}
        \caption{\spork's merge}
    \end{subfigure}
    \caption{The left, base and right revisions of the license header from file merge \casefileheadermerge, along with \spork's merge. \jdime and \automergeptm do not merge comments, and discard file headers completely.}
    \label{fig:res:file_header_merge}
\end{figure}

It is important to note that conflict quantities and sizes alone do not fully
describe the conflict behavior of a merge tool. While in general, merge tools
should strive for as few and as small conflicts as possible, the presence of a
conflict is positive when there is no best conflict handling decision to be
made. Similarly, a smaller conflict is not always easier to interpret, as it
may be small by virtue of failing to include relevant information. We now
illustrate this important point with examples\footnote{Note that the presence
    of \ldots in a source code snippet indicates that it has been truncated to
fit the paper format.}. Each case study is provided with an identifier on the
form \emph{Cx}, where \emph{x} is an integer. This identifier can be used to
find the complete file merge along with all metadata in our online
appendix\footnote{\sporkappendix}.

As a first concrete example, consider the merge of the file header comment in
\autoref{fig:res:file_header_merge}, stemming from file merge
\casefileheadermerge. \spork correctly produces a conflict as the changes
across revisions are incompatible, while both \jdime and \automergeptm simply
discard the file header comment, producing no conflict. In this case, the
presence of a conflict is good, and \spork produces the most informative output
for the developer.

\begin{figure}[t]
\scriptsize
    \begin{subfigure}[t]{0.97\linewidth}
        \begin{tabular}{c}
            \begin{lstlisting}[basicstyle=\scriptsize,language=java]
<<<<<<< LEFT
!transport.isSuccessful()) parseAndThrowException(result);
=======
!transport.isSuccessful()) parseAndThrowException(result, jobInfo.getContentType());
>>>>>>> RIGHT\end{lstlisting}
        \end{tabular}
        \caption{Conflict from \spork's merge of \caseconservative caused by too conservative left/right matching. \jdime and \automergeptm produce the right revision's contribution as the merged output.}
        \label{fig:res:spork_conservative_matching}
    \end{subfigure}
    ~
    \begin{subfigure}[t]{0.97\linewidth}
            \begin{tabular}{c}
                \begin{lstlisting}[linewidth=0.95\textwidth,basicstyle=\scriptsize,language=java,]
<<<<<<< LEFT
long idleThreadKeepAliveMillis = 60000;
=======
private static final String DEFAULT_EXCHANGE_NAME = "";
>>>>>>> RIGHT\end{lstlisting}
        \end{tabular}
        \caption{Conflict between two unrelated and textually far removed
    fields from \jdime's/\automergeptm's merge of \caseaggressive, caused by too aggressive left/right matching. \spork correctly adds
both fields at their respective points of insertion.}
        \label{fig:res:jdime_aggressive_matching}
    \end{subfigure}
    \caption{Snippets showing drawbacks of too conservative and too aggressive left/right matchings}
    \label{fig:res:left_right_matching_balancing}
\end{figure}

\begin{figure*}[t]
\scriptsize
\centering
    \begin{subfigure}[t]{0.59\textwidth}
        \centering
        \begin{tabular}{c}
            \begin{lstlisting}[linewidth=0.95\textwidth,basicstyle=\scriptsize,language=java,numbers=none]
     @Test
     public void testNonNullNativeIgnoreingDocumentationParameterMatcher() {
         context.checking(new Expectations() {{
-            exactly(1).of (mock).withBoolean(with(any(Boolean.class)));
-            exactly(1).of (mock).withByte(with(any(Byte.class)));
...
+            exactly(1).of(mock).withBoolean(with.booleanIs(anything()));
+            exactly(1).of(mock).withByte(with.byteIs(anything()));
...
         }});\end{lstlisting}
        \end{tabular}
        \caption{Base/left line-based diff. Lines preceded by \lstinline{-} and \lstinline{+} indicate deletions and additions, respectively.}
    \end{subfigure}
    ~
    \begin{subfigure}[t]{0.39\textwidth}
        \centering
            \begin{tabular}{c}
                \begin{lstlisting}[linewidth=0.95\textwidth,basicstyle=\scriptsize,language=java,numbers=none]
     @Test
-    public void testNonNullNativeIgnoreing...
+    public void testNonNullNativeIgnoring...\end{lstlisting}
        \end{tabular}
        \caption{Base/right line-based diff. Lines preceded by \lstinline{-} and \lstinline{+} indicate deletions and additions, respectively.}
    \end{subfigure}
    ~
    \begin{subfigure}[t]{.97\textwidth}
        \centering
            \begin{tabular}{c}
                \begin{lstlisting}[linewidth=0.95\textwidth,basicstyle=\scriptsize,language=java,numbers=none]
<<<<<<< LEFT
...
  @Test public void testNonNullNativeIgnoreingDocumentationParameterMatcher() {
    context.checking(new Expectations() {
      {
        exactly(1).of(mock).withBoolean(with.booleanIs(anything()));
...
  }
=======
>>>>>>> RIGHT

...
  @Test public void testNonNullNativeIgnoringDocumentationParameterMatcher() {
    context.checking(new Expectations() {
      {
        exactly(1).of(mock).withBoolean(with(any(Boolean.class)));
...
  }\end{lstlisting}
        \end{tabular}
        \caption{\jdime's/\automergeptm's merge, with the left revision's version of the method in a conflict, followed by the right revision's version of the method outside the conflict hunk}
    \end{subfigure}
    \caption{Line-based base/left and base/right diffs from file merge \caserename. The left revision edits the body of a test method, and the right revision fixes a typo in the method's name. \jdime and \automergeptm do not detect the rename, and produce merge conflict. \spork correctly
    merges the name change in the right revision with the body changes in the left, producing no conflict.}
    \label{fig:res:method_rename}
\end{figure*}

The opposite is also prominent, i.e. that some non-conflicting edits are
detected as conflicts. \autoref{fig:res:left_right_matching_balancing} shows
the effects of too conservative and too aggressive left/right matchings from
file merges \caseconservative and \caseaggressive, respectively. In
\autoref{fig:res:spork_conservative_matching}, \spork's conservative left/right
matching causes it to fail to match near-identical subtrees inserted in the
left and right revisions, thus producing a coarse conflict where the right
revision's part is a strict superset of the left. This can be automatically
resolved, and a reasonable resolution to the conflict is the right revision,
which is what \jdime and \automergeptm produce. However, too aggressive
left/right matching also causes problems with conflicts. In
\autoref{fig:res:jdime_aggressive_matching}, \jdime and \automergeptm match two
completely unrelated fields that are added some 100 lines away from each other
in the left and right revisions, respectively, and therefore produce a
nonsensical conflict. \spork on the other hand inserts the fields
appropriately, without conflict.

We now provide evidence of \spork’s move and update detection capability being
beneficial. \autoref{fig:res:method_rename} shows parts of the base/left and
base/right diffs from file merge \caserename, where the left revision edits the
body of a test method, and the right revision renames said method. \jdime and
\automergeptm both treat the rename in the right revision as a deletion of the
original method, and an insertion of an entirely new method. The deletion
interferes with the edit in the method's body in the left revision This results
in a delete/edit conflict containing the left revision's version of the method.
As the right revision's renamed method is seen as an insertion, it is printed
outside the conflict hunk. Thus, in failing to match the renamed method of the
right revision to the edited method in the left revision, the merge conflict
produced is not only unnecessary, but it also fails to include the right
revision's version of the method in the conflict hunk. Thus, a smaller conflict
hunk is not always easier to understand. \spork on the other hand performs the
merge without conflict, as it detects the right revision's rename as an update
of the method node's content, which is unrelated to the left revisions edits in
its subtree (see the description of \textsc{spork-3dm} in
\autoref{sec:spork:3dm_merge} for the separation of content and structure).

The takeaway of these illustrative case studies is that \spork exhibits
differing and desirable merge properties from \jdime and \automergeptm. This
experiment also recalls that conflict quantity and size are indicative but not
perfect metrics~\cite{cavalcanti2019impact}, as there may be some degenerate
cases. For example, when conflicts occur inside comments or formatting, a merge
tool which does not support comments or formatting preservation may produce
zero conflict while missing an essential part of the merge.

\begin{mdframed}[style=mpdframe]
    \textbf{Answer to RQ1. \spork produces fewer and smaller conflicts than \jdime, and is on par with \automergeptm. All assessed merge tools sometimes produce abnormally large conflicts (\autoref{fig:res:conflict_size_histogram}) but \spork to a lesser extent.}
\end{mdframed}

\begin{figure}[t]
    \centering
    \resizebox{\linewidth}{!}{\begingroup \makeatletter 
\makeatother \endgroup  }
    \caption{Histogram of file merge running times for \spork, \jdime and
    \automergeptm. Lower is better. Each histogram bin contains the frequency
    of values in the range $[L, R)$, where $L$ and $R$ are the values to the left
    and right of the bin, respectively.}
    \label{fig:res:running_times_histogram}
\end{figure}

\subsection{RQ2: Running Time} \label{sec:res:running_times}

The running time of a tool on a given file merge is computed as the median wall
time of 10 executions. We only consider the 1657 file merges where
all of \spork, \jdime and \automergeptm produce a non-empty merged file. It is
noteworthy that there are cases where \jdime and \automergeptm fail due to
timing out at 300 seconds. \jdime suffers 16 timeouts and \automergeptm suffers
7, while \spork exhibits no timeouts. The exclusion of these timeouts is
conservative, as it clearly benefits \jdime and \automergeptm.

In the median case, \spork has a running time of 1.17 seconds, \jdime has a
running time of 1.32 seconds and \automergeptm has a running time of 1.48
seconds. Per this median value, \spork is the fastest out of the three.
\spork being a faster tool is further reinforced by the sum of running times:
\spork's total running time is 2415 seconds, which is 51\% faster than \jdime's
4912 seconds, and 55\% faster than \automergeptm's 5360 seconds.
The histogram of running times in \autoref{fig:res:running_times_histogram}
further exposes performance differences. \jdime and \automergeptm have more of
the smallest running times (leftmost bin), with 52 and 38 running times
respectively that are less than 0.5 seconds, while \spork has none. In terms of
the largest running times (rightmost bin), \jdime and \automergeptm have 245
and 282 running times respectively that are larger than or equal to 4 seconds,
whereas \spork only has 52.  Furthermore, \spork's maximum running time is 11.9
seconds, while \jdime and \automergeptm top out at 287.9 and 287.7 seconds,
respectively. Compared to \jdime, spork is faster in 963 cases and slower in
the remaining 694 cases. Compared to \automergeptm, \spork is faster in 1126
cases and slower in the remaining 531 cases. While \spork is not as fast as
either \jdime or \automergeptm in the best case, it is faster in the median
case, and significantly reduces the amount and magnitudes of excessive running
times larger than 4 seconds.

We use a Friedman test to determine if further analysis of the results is
relevant, with the null hypothesis that the results from the different tools are
the same. The test yields a p-value of 7.88e-247, so we reject the null
hypothesis and proceed with further analyses.

We use a two-sided Wilcoxon signed-rank test to test the following hypothesis:

\begin{quote}
    $H_0^5$: There is no difference between \spork's and \jdime's running times

    $H_a^5$: There is a difference between \spork's and \jdime's running times
\end{quote}

\noindent
The test yields a p-value of 1.80e-54, and we therefore accept the alternative
hypothesis. The effect size RBC is -0.441, indicating that \spork's running times are
smaller than \jdime's.

We use a two-sided Wilcoxon signed-rank test to test the following
hypothesis:

\begin{quote}
    $H_0^6$: There is no difference between \spork's and \automergeptm's running times

    $H_a^6$: There is a difference between \spork's and \automergeptm's running times
\end{quote}

\noindent
The test yields a p-value of 1.74e-121, and we therefore accept the
alternative hypothesis. The effect size RBC is -0.666, indicating that \spork's running
times are smaller than \automergeptm's.

\begin{mdframed}[style=mpdframe]
\textbf{Answer to RQ2. \spork is a faster merge tool than the state of the art. In particular, it has fewer exceedingly long running times which makes it more useful in practice for the developer.} 
\end{mdframed}

\subsection{RQ3: Formatting Preservation}\label{sec:res:formatting}
Formatting preservation is measured as the diff size (the sum of insertions and
deletions in a diff) between the replayed merge produced by the merge tool and
the expected revision committed by the developer, considered as ground truth.
We use two metrics at different levels of granularity: a \linediff as well as a
\chardiff. The results can be interpreted as the amount of lines and the amount
of characters by which the produced and expected revisions differ. We consider
the 1402 file merges in which all of \spork, \jdime and \automergeptm produce
conflict-free merges.

\spork produces file merges with a median \linediff size of 65, which
represents a 78\% reduction compared to \jdime's median of 308.5, and a
79\% reduction compared to \automergeptm's median of 314.5. This clearly
shows that \spork preserves more formatting than the other tools.  The
histogram in \autoref{fig:res:gitdiff_size_histogram} shows \spork's clear
advantage over \jdime and \automergeptm. Compared to \jdime, \spork
produces smaller \linediff sizes for 1336 cases, of equal size in 3 cases,
and larger ones in the remaining 63.  Compared to \automergeptm, \spork
produces smaller \linediff sizes in 1341 cases, of equal size in 3 cases, and
larger ones in the remaining 58.

\begin{figure}
    \centering
    \resizebox{\linewidth}{!}{\begingroup \makeatletter 
\makeatother \endgroup  }
    \caption{Histogram of \linediff sizes for \spork, \jdime and
    \automergeptm. Lower is better. Each histogram bin contains the frequency
    of values in the range $[L, R)$, where $L$ and $R$ are the
    values to the left and right of the bin, respectively.}
    \label{fig:res:gitdiff_size_histogram}
\end{figure}

The trend set in the \linediff comparison carries over to the \chardiff
measurements. \spork produces file merges with a median \chardiff size of 528,
which represents a 75\% reduction compared to \jdime's median of 2181, and a
78\% reduction compared to \automergeptm's median of 2430. The histogram in
\autoref{fig:res:chardiff_size_histogram} shows \spork's clear advantage over
\jdime and \automergeptm. Compared to \jdime, \spork produces smaller character
diff sizes in 1286 cases, and larger ones in the remaining 116.  Compared to
\automergeptm, \spork produces smaller \chardiff sizes in 1301 cases, and
larger ones in the remaining 101. These numbers correspond well with the
line-based \linediff, indicating that it is a good approximation for the
overall textual similarity of two files.

\begin{figure}[t]
    \centering
    \resizebox{\linewidth}{!}{\begingroup \makeatletter 
\makeatother \endgroup  }
    \caption{Histogram of \chardiff sizes for \spork, \jdime and
    \automergeptm. Lower is better. Each histogram bin contains the frequency
    of values in the range $[L, R)$, where $L$ and $R$ are the
    values to the left and right of the bin, respectively.}
    \label{fig:res:chardiff_size_histogram}
\end{figure}

To illustrate \spork's improvements, we present a final case study.
\autoref{fig:res:reprinting_comparison} shows a complex conditional expression
from file merge \caseformatting. The condition of the if-statement is complex
both with respect to the number of clauses and with respect to formatting (lots
of ad hoc indentation and line breaks). Through high-fidelity pretty-printing
of the method containing this if-statement, \spork precisely reproduces said
formatting. In contrast, \jdime's pretty-printer both changes the indentation
from 4 spaces to 2, and collapses the entire first condition into a single line
of 280 characters, completely ruining readability. This also applies to
\automergeptm by virtue of using \jdime's pretty-printer.

Our manual analysis confirms that the small diff sizes for \spork's
merges can be attributed to the \spork's high-fidelity pretty-printing
that preserves the original indentation, style and formatting, as
explained in \autoref{sec:spork:pretty_printing}. \spork is
able to copy the original source code of certain elements involved in a merge and
print it as-is into the output file. This is in contrast to \jdime and
\automergeptm, which only perform low-fidelity pretty-printing with its
own formatting style.

We use a Friedman test to determine if further analysis of the
\linediff sizes is relevant, with the null hypothesis that the
results from the different tools are the same. The test yields a p-value of
0 with machine precision, so we reject the null hypothesis and proceed with
further analyses.

\begin{figure}[t]
\scriptsize
\centering
    \begin{subfigure}[t]{.97\linewidth}
        \centering
            \begin{tabular}{c}
                \begin{lstlisting}[language=java]
if (parentContext != null
        && parentContext.object != null
        && ("java.util.ArrayList".equals(parentName)
        || "java.util.List".equals(parentName)
        || "java.util.Collection".equals(parentName)
        || "java.util.Map".equals(parentName)
        || "java.util.HashMap".equals(parentName))) {
    parentName = parentContext.object.getClass().getName();
    if (parentName.equals(parentClassName)) {
        param = parentContext.object;
    }
}\end{lstlisting}
        \end{tabular}
        \caption{\spork's output, identical to the developer merge}
        \vspace{2em}\end{subfigure}
    ~
    \begin{subfigure}[t]{.97\linewidth}
        \centering
            \begin{tabular}{c}
                \begin{lstlisting}[language=java]
if (parentContext != null && parentContenxt...
  parentName = parentContext.object.getClass().getName();
  if (parentName.equals(parentClassName)) {
    param = parentContext.object;
  }
}\end{lstlisting}
        \end{tabular}
        \caption{\jdime's/\automergeptm's output. The entire condition has been written out on a single 280 characters long line (note truncation: \ldots), which would not be acceptable for the developer.}
    \end{subfigure}
    \caption{Comparison between \spork' and \jdime's formatting preservation on part of file merge \caseformatting}
    \label{fig:res:reprinting_comparison}
\end{figure}

We use a two-sided Wilcoxon signed-rank test to test the following hypothesis:

\begin{quote}
    $H_0^7$: There is no difference between the \linediff sizes of file
    merges produced by \spork and \jdime

    $H_a^7$: There is a difference between the \linediff sizes of file
    merges produced by \spork and \jdime
\end{quote}

\noindent
The test yields a p-value of 1.85e-213, and we therefore accept the alternative
hypothesis that there is a difference between the \linediff sizes of
merges produced by the tools. The RBC is -0.963, indicating that \spork produces
merges with lesser \linediff sizes than \jdime.

We use a two-sided Wilcoxon signed-rank test to test the following
hypothesis:

\begin{quote}
    $H_0^8$: There is no difference between the \linediff sizes of file
    merges produced by \spork and \automergeptm

    $H_a^8$: There is a difference between the \linediff sizes of file
    merges produced by \spork and \automergeptm
\end{quote}

\noindent
The test yields a p-value of 1.85e-213, and we therefore accept the alternative
hypothesis that there is a difference between the \linediff sizes of
merges produced by the tools. The RBC is -0.963, indicating that \spork produces
merges with lesser \linediff sizes than \automergeptm.

We use a Friedman test to determine if further analysis of the
\chardiff sizes is relevant, with the null hypothesis that the
results from the different tools are the same. The test yields a p-value of
0 with machine precision, so we reject the null hypothesis and proceed with
further analyses.

We use a two-sided Wilcoxon signed-rank test to test the following hypothesis:

\begin{quote}
    $H_0^9$: There is no difference between the \chardiff sizes of file
    merges produced by \spork and \jdime

    $H_a^9$: There is a difference between the \chardiff sizes of file
    merges produced by \spork and \jdime
\end{quote}

\noindent
The test yields a p-value of 1.29e-199, and we therefore accept the alternative
hypothesis that there is a difference between the \chardiff sizes of
merges produced by the tools. The RBC is -0.929, indicating that \spork produces
merges with lesser \chardiff than \jdime.

We use a two-sided Wilcoxon signed-rank test to test the following
hypothesis:

\begin{quote}
    $H_0^10$: There is no difference between the \chardiff of file
    merges produced by \spork and \automergeptm

    $H_a^10$: There is a difference between the \chardiff of file
    merges produced by \spork and \automergeptm
\end{quote}

\noindent
The test yields a p-value of 2.26e-205, and we therefore accept the alternative
hypothesis that there is a difference between the character sizes of
merges produced by the tools. The RBC is -0.944, indicating that \spork produces
merges with lesser \linediff sizes than \automergeptm.

\begin{mdframed}[style=mpdframe]
    \textbf{Answer to RQ3. \spork preserves formatting to a greater extent
    than \jdime and \automergeptm. \spork produces smaller line and character diffs
    in more than 90\% of cases with median diff size reductions of 75\% and above.}
\end{mdframed}

\subsection{Recapitulation}

In our experiments, we have answered three research questions
targeting different facets of structured merge: conflicts, running
times and formatting preservation. We have systematically and
quantitatively compared our contribution, \spork, against the relevant
state-of-the-art, \jdime and \automergeptm. We summarize the quantitative
results in \autoref{tab:res:summary}. Regarding conflicts (RQ1), \spork
performs better than \jdime and on par with \automergeptm. Regarding running
times (RQ2), \spork is slightly faster in the median case, but more
importantly reduces both amounts and magnitudes of excessive running times.
Regarding formatting preservation, which is our main contribution, \spork
decreases the formatting changes by an order of magnitude. According to this
evaluation, \spork can be considered to be pushing the state of the art of
software merging.

\begin{table}[t]
  \renewcommand{\arraystretch}{1.7}
  \centering
  \begin{tabular}{l|p{3cm}rrrr}
                                                                     &                                 & \spork & \jdime & \textsc{aptm} \\
    \hline
    \multirow{5}{1.5cm}{\textbf{RQ1: Conflicts}}                     & \# considered merges            & 255    & 255    & 255 \\
                                                                     & \# files with conflicts         & 125    & 191    & 145 \\
                                                                     & \# conflict hunks               & 227    & 376    & 245 \\
                                                                     & \# conflicting lines (total)    & 2446   & 13975  & 6635 \\
                                                                     & \# conflict sizes $\geq$ 20 LOC & 24     & 52     & 45 \\
    \hline
    \multirow{5}{1.5cm}{\textbf{RQ2: Running times}}                 & \# considered merges            & 1667   & 1667   & 1667 \\
                                                                     & median running time             & 1.18s  & 1.32s  & 1.48s \\
                                                                     & total running time              & 2435s  & 4937s  & 5388s \\
                                                                     & \# running times $<$ 0.5s       & 0      & 52     & 38 \\
                                                                     & \# running times $\geq$ 4s      & 53     & 248    & 285 \\
    \hline
    \multirow{3}{1.5cm}{\textbf{RQ3: Formatting}}                    & \# considered merges            & 1402   & 1402   & 1402 \\
                                                                     & median \linediff size           & 65     & 308.5  & 314.5 \\
                                                                     & median char diff size           & 528    & 2181   & 2430 \\

    \hline
  \end{tabular}
  \caption{Summary of our quantitative results.}
  \label{tab:res:summary}
\end{table}

\section{Discussion}\label{sec:discussion}

The results of our experiments indicate that \spork performs well overall.
In this section, we discuss the limitations we identified, as well as the threats to the validity of our experiment.

\subsection{Limitations of \spork}
\label{sec:disc:limitations_spork}

\spork has two limitations when it comes to handling conflicts. The first one
is the problem with move and delete conflicts, which are currently handled with
textual representations of the subtrees involved. Move conflicts in particular
are difficult to handle, and pose a problem that is introduced solely due to
\spork being move-enabled. While there are file merges in the results that
\spork can merge due to being move-enabled, such as method renaming, it is
unclear whether the benefits outweigh the drawbacks. Therefore, a future study
to compare move-enabled merge to non-move-enabled merge is called for.

The second conflict-related limitation is that \spork ignores so-called
delete/edit conflicts, which occur when one revision deletes a subtree where
the other revision performs edits. In \textsc{3dm-merge}, such a deletion
silently overrides any edits in the subtree~\cite{Lindholm2004}, and \spork has
no additional measure in place to detect such conflicts. This limitation is
thus inherited from \textsc{3dm-merge}. While detecting a delete/edit conflict
in \textsc{3dm-merge} is possible through post-processing of the change set~\cite{Lindholm2004},
finding the correct way to represent it in the merged AST is less straightforward
and requires non-trivial extensions of \spork. Combined with the findings
presented in the case studies in \autoref{sec:res:conflicts}, more in-depth
analysis of conflict behavior along the lines of those conducted by Cavalcanti
et al.~\cite{cavalcanti2019impact} and Tavares et al. \cite{Tavares2019} is
therefore necessary to draw accurate conclusions about conflict handling.

Furthermore, there are limitations in \spork's high-fidelity
pretty-printing, which is a fundamentally hard problem \cite{Waddington2007}.
While high-fidelity pretty-printing is one of \spork's primary advantages over
the other structured merge tools, it is not perfect. In the current
implementation, high-fidelity pretty-printing is only enabled for type members
and comments that stem from a single revision. More granular elements are
printed with low-fidelity pretty-printing. This often causes \spork to alter
formatting in undesirable ways, such as by printing redundant parentheses not
present in the original source code~\cite{Adzemovic2020}, or by failing to
reproduce ad-hoc indentation like \jdime does in
\autoref{fig:res:reprinting_comparison}. To sum up, while \spork greatly
improves over the related work with respect to formatting and readability of
merges, the difficulty of the problem calls for future research and engineering
about formatting preservation.

\spork also exhibited 34 crashes in the experiments, indicating
unhandled corner cases. It should be noted that 15 of these errors were caused
by parse errors in \textsc{spoon}, and were thus completely outside of \spork's
control.

\subsection{Threats to Validity}
\label{sec:disc:threats}

The primary threats to external validity are the diversity and
representativeness of the dataset, as defined by Nagappan et al.~\cite{Nagappan2013}.
The diversity of the dataset is a critical aspect enabling the results to
generalize. Our dataset consists of open-source \textsc{java} projects from the
\textsc{github} platform, which means that the results do not necessarily
generalize to other platforms or closed-source projects. Discarding  projects
and merge scenarios that failed to build with \textsc{maven} also limits the
diversity of the dataset, both by honing in on projects using \textsc{maven}
and by enforcing that the projects build.

A threat to representativeness is the fact that our methodology can only
discover merge scenarios that are explicitly present in the commit history,
which notably excludes merges that have been squashed or occurred during
rebasing~\cite{Bird2009,Ji2020a}. Furthermore, as no project was allowed to
contribute more than 15 merge scenarios, the dataset is not representative of
the population of merge scenarios in terms of proportions. This is however
necessary, as trial runs of the experiments without this restriction had a few
of the largest projects completely determine the outcome. By limiting the
amount of merge scenarios per project, smaller projects with fewer merge
scenarios are also allowed to meaningfully impact the results. This makes the
results more representative of the population of projects rather than the
population of merge scenarios.

There are three primary threats to internal validity, all of which are related
to the execution of the experiments. First, running time measurements are
not perfectly reliable because of the underlying variance of the system, even
with 10 repetitions of each merge. Second, the experiment scripts are
relatively complex, and there is a possibility that they contain errors. To
mitigate such risks, all our benchmark scripts are made publicly available
in our online appendix\footnote{\sporkappendix}.
Third, the results are only valid for one set of tuning parameters, and these
are not necessarily optimal for any of the tested tools. Notably, the
experiments were executed with \textsc{jdime}'s default settings. This for
example means that its lookahead heuristics for identifying renamed methods and
shifted code were not enabled, which if enabled could have helped avoid some
conflicts at the cost of increased running time~\cite{LeBenich2017}.

The 83 file merges excluded on the basis of at least one tool
exhibiting a merge failure also pose a threat to validity. As the overlap in
failing file merges is small between the tools, there is a possibility that
these exclusions are more advantageous for some tools than others. For example,
excluding a file merge where tool \textsc{A} times out benefits the running
time results of tool \textsc{A}. Similarly, excluding a file merge where tool
\textsc{B} crashes or produces an empty file can mask poor formatting
preservation or large amounts and sizes of conflicts, potentially benefiting
tool \textsc{B}.

\section{Related Work}\label{sec:related_work}

Merging of source code is an active research field. This section presents the
most closely related work on merge tools in \autoref{sec:improved_tools}, and
other approaches to assist in the merging of code in
\autoref{sec:rw:other_approaches}.

\subsection{Structured and semistructured merge}\label{sec:improved_tools}
This section outlines related work on structured and semistructured merging.
\autoref{sec:structured_diff} presents structured diff algorithms,
\autoref{sec:structured_merge} presents complete structured merge tools and
\autoref{sec:semistructured_merge} presents related work on semistructured
merge.

\subsubsection{Structured diff algorithms}\label{sec:structured_diff}
The distinction between an unstructured diff algorithm and a structured
one is that the former operates on raw text, while the latter operates
on some form of structure that the text encodes~\cite{Mens2002}. Most often,
that entails some form of tree structure, ranging from ordered trees to
represent structured text documents~\cite{Chawathe1996} to fully resolved
ASTs~\cite{Falleri2014}. More generalized graph representations can also
be utilized~\cite{Apiwattanaponga,Mens2002}.

\textsc{ladiff} represents one of the earliest structured diff algorithms
that can deal with insertions, deletions, updates and
moves~\cite{Chawathe1996}. It targets structured text documents, such as LaTeX
and HTML. The algorithm relies heavily on an assumption that each leaf node in
a tree $T_1$ has at most one highly similar leaf node in another tree $T_2$.
This makes it unsuitable for source code differencing.

\textsc{changedistiller} improves upon \textsc{ladiff} by removing the
assumption of unique matchings for leaf nodes~\cite{Fluri2007}, making it
more suitable for source code differencing. Leaf nodes are however represented
as text, meaning that there is still room for increased granularity.

\textsc{gumtree} is a structured diff algorithm that like \textsc{ladiff} and
\textsc{changedistiller} can operate on insertions, deletions, updates and
moves~\cite{Falleri2014}. However, it operates on a fully resolved AST, making
it more granular. We make use of \textsc{gumtree} in our own work.

\textsc{calcdiff} is another structured diff algorithm that operates
on a control flow graph instead of an AST~\cite{Apiwattanaponga}. It is
specifically designed to target object-oriented languages, and in particular
with static code analysis in mind, such as being able to predict test coverage
changes based on changes to the production source code.

\subsubsection{Structured merge tools}\label{sec:structured_merge}
Structured merge tools typically make use of a structured diff algorithm
to identify changes across revisions, and based on that information use
varying strategies for computing a merge. The topic was first studied
in the early 1990s~\cite{Westfechtel1991}.

\textsc{jdime} is a three-way structured merge tool for \textsc{java} that
implements its own tree differencing and merging
algorithms~\cite{Lessenich2012,Le_enich_2014}. The matching step is simplistic
and can only detect insertions and deletions. A heuristic lookahead mechanism
built on top of the matching does however allow for limited move and update
detection~\cite{LeBenich2017}. The work on \jdime is closely related to
our own work, and we have drawn a great deal of inspiration from it. What
sets our work apart is more powerful tree matching, a focus on providing
minimal textual diffs with high-fidelity pretty-printing, and overall more
modern components allowing support for newer versions of \textsc{java}.

Another approach to structured merge is to use a generic, textual
representation of ASTs, and then merge with a standard line-based merge
algorithm\cite{Asenov2017}. The proposed algorithm can work either with unique
identifiers stored across revisions to avoid the need for tree differencing, or
use a differencing algorithm such as \textsc{gumtree} to compute matchings.

\textsc{3dm} is a move-enabled three-way merge tool designed for XML documents,
with a novel merge algorithm that is applicable to any form of ordered
tree~\cite{Lindholm2004}. It operates on units of small node contexts of three
nodes; a parent node, and two of its children in the order they appear in its
child list. This makes the merge granular, and it is also efficient with a time
complexity of $\mathcal{O}(n*log(n))$. We implement the merge algorithm
from \textsc{3dm} in our own work.

Another approach for merging XML documents is to apply diffs computed on one
version of a document to another version of it~\cite{Roennau2008,Roennau2009}.
This approach has the benefit of not requiring all three revisions to be
present on the same machine, which may prove useful in situations where
bandwidth is highly limited. It is however by nature less precise than a
traditional three-way merge, such as the one implemented by \textsc{3dm}.

\subsubsection{Semistructured merge tools}\label{sec:semistructured_merge}
Semistructured merge tools represent an attempt to find a middle-ground between
structured and unstructured merging in terms of accuracy and running time
performance~\cite{Apel2011}. The idea is to  merge high-level elements such as
method headers structurally, and use unstructured merge within fine-grained
code elements such as method bodies.

\textsc{fstmerge} is the earliest example of semistructured
merge~\cite{apel2010semistructured}, and provides a framework for implementing
semistructured merge tools. Merge tools built on \textsc{fstmerge} have been
shown to improve upon unstructured merge for \textsc{java}, \textsc{python} and
\textsc{c\#}~\cite{Apel2011,Cavalcanti2015,Cavalcanti2017}. An implementation
for \textsc{javascript} also exists, but the approach of semistructured merge
yields significantly smaller improvements for \textsc{javascript} than it does
for a language like \textsc{java}~\cite{Tavares2019}.

\textsc{intellimerge} presents a different approach to semistructured merge for
\textsc{java}~\cite{Shen2019}. It uses a lightweight graph to represent the
overall structure of a program, while keeping method bodies in textual form.
While graph-based merging techniques typically suffer from excessive running
times~\cite{Falleri2014,Shen2019}, \textsc{intellimerge} is shown to be even
faster than a comparable specialization of \textsc{fstmerge}.

\subsection{Other Approaches}\label{sec:rw:other_approaches}
Orthogonally to the development of better merge tools, there are two other
major approaches to assisting the merging of code. The first of these is
\emph{conflict resolution helpers}. The most straightforward of such tools are
simple visualizers of conflicts, such as
\textsc{kdiff3},
\textsc{meld} and
\textsc{winmerge}. More involved tools may
provide collaborative online environments for solving
conflicts~\cite{Nieminen2012}, automated suggestions for which developers are
best equipped to solve some given conflict~\cite{Costa2016}, replaying of
individual edits~\cite{Nishimura2016} and even synthesizing of solutions to
conflicts~\cite{Zhu2018}.

The second major approach is to avoid conflicts by predicting them before they
occur. Workspace awareness tools such as \textsc{syde}~\cite{Hattori2010},
\textsc{palantir}~\cite{Sarma2012}, \textsc{cassandra}~\cite{kasi2013cassandra}
and \textsc{crystal}~\cite{Brun2013} monitor the workspaces of individual
developers and try to predict where conflicts may occur with other developers.
This is typically done by preemptively merging developers' branches with each
other, with some variations in the exact mechanisms, the merge tools used and
the amount of validation of the merged systems. A more recent trend is to do
lightweight feature analysis in order to predict
conflicts~\cite{Accioly2018,Lesenich2017,Dias2020,Owhadi-Kareshk2019a}, or
predict the difficulty of resolving a conflict that has already
manifested~\cite{Brindescu2020}. This can potentially enhance workspace
awareness tool accuracy while also reducing computational cost.

\section{Conclusion}\label{sec:conclusions}

In this paper, we have presented a novel structured merge system for
\textsc{java}, called \spork. \spork, uniquely based on the \textsc{3dm}
algorithm, embeds essential domain knowledge of the \textsc{java} programming
language in order to minimize the amount of conflicts and the impact on
formatting. We have presented a systematic and large scale empirical
evaluation, showing that \spork makes significant improvements to key metrics
of merging, including running times and preservation of source code formatting.

We observe that formatting is an important aspect of source code that
developers care deeply about, and plays a prominent role in readability and
maintenance. As such, merge tools that do not preserve the formatting that
developers have put in place are unlikely to be widely adopted. While \spork
presents a major improvement over comparable tools in terms of preserving
formatting, it still in part makes use of low-fidelity pretty-printing that
alters formatting. We believe that future research on structured merge should
focus on improving formatting preservation even further, as without near
perfect preservation of formatting, real-world applicability of structured
merge remains limited.

\balance

\end{document}